%% file: main.tex
\documentclass[fleqn, usenatbib]{mnras}
\usepackage{amsmath}
\usepackage{newtxtext,newtxmath}

\usepackage[utf8]{inputenc}
\usepackage[T1]{fontenc}
\usepackage{ae,aecompl}

\usepackage{graphicx}
\usepackage{aas_macros}

\usepackage[figuresright]{rotating}
\usepackage{booktabs}
\usepackage[english]{babel}
\usepackage{microtype}
\usepackage[pdftex]{lscape}
\usepackage[usenames,dvipsnames]{xcolor}
\usepackage{xspace}
\usepackage{my_def}
\usepackage{float}
\DeclareRobustCommand{\VAN}[3]{#2}
\let\VANthebibliography\thebibliography
\def\thebibliography{\DeclareRobustCommand{\VAN}[3]{##3}\VANthebibliography}

\DeclareRobustCommand{\DE}[3]{#2}
\let\DEthebibliography\thebibliography
\def\thebibliography{\DeclareRobustCommand{\DE}[3]{##3}\DEthebibliography}

\newcommand*{\phib}{\phi_\mathrm{b}}
\newcommand*{\phif}{\phi_\mathrm{f}}
\newcommand*{\alphab}{\alpha_\mathrm{b}}
\newcommand*{\alphaf}{\alpha_\mathrm{f}}

\newcommand*{\EAGLE}{\textsc{eagle}\xspace}
\newcommand*{\CEAGLE}{\textsc{c-eagle}\xspace}

\bibpunct{(}{)}{;}{a}{}{,}

\date{Draft, \today}

\title{The luminosity of cluster galaxies in the Cluster-EAGLE simulations}

\author[A.~Negri, C.~Dalla Vecchia, J.~A.~L.~Aguerri, Y.~Bahé]
{Andrea Negri,$^{1,2}$\thanks{E-mail: anegri@iac.es (AN)}
Claudio Dalla Vecchia,$^{1,2}$
J. Alfonso L. Aguerri,$^{1,2}$
Yannick Bahé$^3$
\\
$^1$Instituto de Astrof\'isica de Canarias, C/V\'ia L\'actea s/n, E-38205 La Laguna, Tenerife, Spain  \\
$^2$Departamento de Astrof\'isica, Universidad de La Laguna, Av.~Astrof\'isico Francisco S\'anchez s/n, E-38206 La Laguna, Tenerife, Spain \\
$^3$Leiden Observatory, Leiden University, PO Box 9513, NL-2300 RA Leiden, The Netherlands \\
}

\date{Accepted XXX. Received YYY; in original form ZZZ}
\pubyear{2021}
\begin{document}
\label{firstpage}
\pagerange{\pageref{firstpage}--\pageref{lastpage}}
\maketitle

\begin{abstract}
We computed the luminosity of simulated galaxies of the \CEAGLE project, a suite of 30 high-resolution zoom-in simulations of galaxy clusters based on the \EAGLE simulation. The AB magnitudes are derived for different spectral bands, from ultraviolet to infrared, using the simple stellar population modeling based on the E-MILES stellar spectra library. We take into account obscuration due to dust in star forming regions and diffuse interstellar medium. The $g-r$ colour-stellar mass diagram, at $z=0.1$, presents a defined red sequence, reaching $g-r \simeq 0.8$, 0.05~dex redder than \EAGLE at high masses, and a well populated blue cloud, when field galaxies are included. The clusters' inner regions are dominated by red-sequence galaxies at all masses, although a non-negligible amount of blue galaxies are still present. We adopt Bayesian inference to compute the clusters LFs, testing for statistical significance of both single and double Schechter functions. The multicolour LFs at $z=0$ show a knee luminosity that peaks in the infrared and increases with the cluster's mass. The faint-end is weakly dependent on colour and mass and shows an upturn in the optical, bounded between $-1.25$ and $-1.39$, just moderately steeper than the field. The simulations reproduce, within the observational errors, the spectroscopic LFs of the Hercules and Abell 85 clusters, including their faint end upturn. \CEAGLE LFs are in broad agreement with observed LFs taken from SDSS and XXL surveys, up to $z=0.67$, showing a rather flat faint end when the observational constrains are taken into account.
\end{abstract}


\section{Introduction}

Galaxy clusters are the most massive collapsed structures in the Universe, containing hundreds to thousands galaxies, making them one of the largest astrophysical laboratories where galaxy formation and evolution can be studied. In comparison with their isolated counterparts, galaxies in clusters have a lower rate of star formation \citep{Kauffmann+2004,Blanton+2005,Peng+2010,Wetzel+2012,Woo+2013} and little cold gas content \citep{Giovanelli.Haynes1985,Fabello+2012,Hess.Wilcots2013,Brown+2017}. Clusters of galaxies show a strong morphological segregation at low redshift, containing a rich population of early-type galaxies at their center \citep{Visvanathan.Sandage1977,Dressler1980,Bower+1992,DeLucia+2004,Lerchster+2011,vanderBurg+2015,Cerulo+2016,Socolovsky+2018,Zhang+2019,Seth.Raychaudhury2020}.

An important tool to understand galaxy evolution is the luminosity function (LF), which expresses the galaxy number density as a function of luminosity. Usually modelled with a \citet{Schechter1976} function, it is a robust observable, extensively used in both field and cluster environments, and it encodes fundamental information about the different physical properties regulating galaxy evolution \citep{Loveday+1992,Blanton+2003,Loveday+2012,Lan+2016}. Thus, the LF and its evolution is one of the most fundamental observational constrains that theoretical models and numerical simulations must reproduce. 

Claims that the LF depends on the environment and redshift are numerous \citep{DePropris+2003,Hansen+2005,Gilbank+2008,Hansen+2009,Zandivarez.Martinez2011,DePropris+2013,Muzzin+2013,Lan+2016}. The LF bright-end has been well characterized in a variety of environments from field to dense clusters \citep[e.g.][]{Blanton+2005,Bowler+2014,Ricci+2018}. On the other hand, the faint-end dependence on the environment is still a matter of debate. The $\Lambda$ cold dark matter ($\Lambda$CDM) model predicts a subhalo mass function with a faint-end slope $\alpha \simeq -1.9$ \citep{Springel+2008}, steeper than what it is observed for field galaxies \citep[$\simeq -1.1$ to $-1.5$,][]{Blanton+2005}.
Some authors have found that, around $M_r \simeq -18$, the cluster LFs show a faint-end upturn in both stacked and individual clusters \citep{Baldry+2008,Agulli+2014,Tomczak+2014,Moretti+2015,Lan+2016,Agulli+2016,Agulli+2016a,DePropris2017} up to $\simeq -1.6$ (usually modelled with a Schechter function plus an additional power-law or a double Schechter), and in some extreme cases as steep as the halo mass function \citep{Popesso+2006,Drory+2009,DePropris+2018}. However, other work has shown a LF that is either flat or consistent with the field, with no sign of a faint-end upturn \citep{DePropris+2003,Li.White2009,deFilippis+2011,DePropris+2013,Muzzin+2013,Agulli+2017,Ricci+2018,OMill+2019,Aguerri+2020}, or of a faint-end redshift evolution \citep{Andreon2006,DePropris+2007,Muzzin+2007,DePropris+2013,Cerulo+2016,DePropris+2016,Sarron+2018,Puddu+2021,Zenteno+2020}. Finally, the study of the LF faint-end is hampered by the observational magnitude limits, and by the mass-redshift degeneracy, due to the fact that at high redshift only the most massive clusters are observed, making difficult to investigate the LF mass dependence \citep{Martinet+2015,Martinet+2017,Sarron+2018}.

In the last 15 years, cosmological, hydrodynamic simulations have become a fundamental tool in the understanding of the physical processes that shape the evolution of galaxies \citep{Naab.Ostriker2017}.
In this work we adopt the Cluster-\EAGLE (hereafter, \CEAGLE) simulation suite \citep{Bahe+2017,Barnes+2017}, where 30 galaxy clusters, uniformly distributed in virial mass within the range\footnote{$M_{200}$ denotes the total mass contained in a sphere centered on the minimum of the cluster gravitational potential, having a radius $r_{200}$ such as the mean density equals 200 time the critical density of the Universe.} $14 < \log(M_{200}/\Msun) < 15.4$, are simulated with the same resolution of the \EAGLE Ref run (see Section~\ref{sec_CEAGLE} for details). The re-simulated volumes have a radius of $5\times r_{200}$ for all the clusters, whilst a subset of 24 of them are simulated up to $10\times r_{200}$. Simulating such a large volume around a cluster of galaxies allows for the study of a variety of environments within the same overdense region of the Universe: the virial overdensity (the cluster itself), rich of heavily evolved, early-type galaxies; the immediate outskirts (the infall region), populated by both relatively gas-rich galaxies, falling into the cluster for the first time, and splash-back galaxies, completing their first orbit around the cluster centre; the far outskirts, where the large-scale structure emerges with filaments (and the galaxies within) and smaller groups of galaxies.

One main output of numerical simulations is the stellar mass of galaxies, from which the galaxy stellar mass function can be immediately derived \citep[see][for a study of the galaxy stellar mass function in \CEAGLE]{Bahe+2017,Ahad+2021}. However, the main observable is the stellar light, which is converted into stellar mass with empirically calibrated relations between stellar mass and colours of galaxies, or, in the case of multi-band or spectroscopic observations, by matching the spectral energy distribution with stellar population models. In this work, we derive the luminosity of galaxies to be compared with observations. We post-process the \CEAGLE simulation by generating stellar spectra and magnitudes for every stellar particle and galaxy in the simulations, at all output redshifts. We also account for dust attenuation on the stellar light, as function of gas metallicity.

This paper is the first of a series dedicated to the study of LFs in simulated galaxy clusters. In section~\ref{sec_CEAGLE}, we provide a brief summary of the \CEAGLE simulation project. In section~\ref{sec_photometry}, we present the post-processing method employed to compute the luminosity of the simulated galaxies, while in section~\ref{sec_3} we introduce the galaxy selection procedure and the Bayesian method adopted in the calculation of the LFs. As validation of the model, we also compare our results to the luminosity of \EAGLE galaxies \citep{Trayford+2015}. In section~\ref{sec_results}, we calculate the galaxy colour-stellar mass relation for the entire \CEAGLE sample, and we discuss differences and similarities with the red sequence of \EAGLE galaxies. We compute the \CEAGLE stacked LFs at $z=0$, in different bands. from ultraviolet to mid-infrared, and we study the correlations between the LFs, and in particular their faint-end slope, and clusters' mass and colour. Finally, we compare the simulated LFs with observations of individual, nearby clusters and observational surveys at intermediate redshift. We summarise our work in section~\ref{sec_conclusions}. We leave the predictions on the evolution of the cluster galaxies LF to a companion paper.

\section{The Cluster-EAGLE simulation project} \label{sec_CEAGLE}
We briefly describe here the simulations we used in this work. For a full description of the \CEAGLE simulations, their initial conditions and sub-grid physics, see \citet{Bahe+2017} and \citet{Barnes+2017}.

The \CEAGLE simulations consist of 30 zoom-in simulations, each one centred on a massive galaxy cluster at $z=0$. The parent simulation is a dark-matter-only run of volume $(3.2~\Gpc)^3$, from which 30 dark matter haloes have been selected, uniformly distributed in the mass range $14 < \log(M_{200}/\Msun) < 15.4$. In addition, the requirement of not having a more massive neighbouring halo within $20\times r_{200}$ has been applied. The clusters have been simulated up to a radius of $10\times r_{200}$ for 24 of them, forming the \textit{Hydrangea} sample \citep{Bahe+2017}, while 6 of them are simulated up to $5\times r_{200}$.

The \CEAGLE simulations have been performed with the \EAGLE model for galaxy formation  \citep{Schaye+2015,Crain+2015}, with the same spatial and mass resolution of the \EAGLE $(100~\mathrm{Mpc})^3$ Ref-L100 simulation: $m_\mathrm{gas} \simeq 1.8\times 10^6~\Msun$; $m_\mathrm{DM} \simeq 9.7\times 10^6~\Msun$; gravitational softening length of $2.66$ comoving $\mathrm{kpc}$ ($\mathrm{ckpc}$) until $z=2.8$, and $0.7$ proper $\mathrm{kpc}$ ($\mathrm{pkpc}$) at lower redshift. The code is a modified version of the $N$-body Tree-PM SPH code \textsc{p-gadget3} \citep{Springel2005}, with an updated hydrodynamic scheme named \textsc{anarchy}  \citep{Schaye+2015,Schaller+2015}. It includes the pressure-entropy formulation of SPH hydrodynamics \citep{Hopkins2013}, with artificial viscosity switch \citep{Cullen.Dehnen2010}, artificial conductivity switch \citep{Price2008} and the time-step limiter introduced in \citet{Durier.DallaVecchia2012}.
The implemented physics includes radiative cooling, stellar feedback, star formation and the seeding and feedback of black holes \citep{Schaye.DallaVecchia2008,Wiersma+2009,DallaVecchia.Schaye2012,Rosas-Guevara+2015,Schaye+2015}. The \EAGLE simulation consists of different runs, where the stellar and black hole feedback have been calibrated to reproduce a set of observables. In particular, simulations Ref and AGNdT9 differ in the heating temperature applied to a gas particle during black hole feedback, which is $10^{8.5}$ and $10^{9}~\mathrm{K}$ respectively, and in the effective viscosity of the subgrid accretion disk around the black hole, which is 100 times larger in AGNdT9 \citep[see][]{Crain+2015,Schaye+2015}. Since this model performs better when the X-ray luminosity of simulated groups of galaxies are compared observations, it has been adopted in the \CEAGLE simulations.
 
 The employed cosmological model is the standard $\Lambda$CDM with parameters from the \textit{Planck 2013} data combined with baryonic acoustic oscillations, WMAP polarization and high multipole moments experiments \citep{PlanckCollaboration+2014a}: $\Omega_\mathrm{b}=0.04825$, $\Omega_\mathrm{m}=0.307$,  $\Omega_\Lambda=0.693$, $h\equiv H_0 / (100~\mathrm{km\,s^{-1}\,Mpc^{-1}})=0.6777$.

 \subsection{Galaxy identification}
 The main output of the simulation consists of 30 full snapshots between $z=14$ and $z=0$, 28 of which are equally spaced in time ($\Delta t = 500~\Myr$) with two additional snapshots at $z=0.366$ and 0.101, to ease the comparison with \EAGLE outputs. In addition, a larger number of reduced-size snapshots, or ``snipshots,'' storing only the most rapidly time-varying quantities, have been saved with a time resolution of $125~\Myr$, boosted at $\Delta t = 25~\Myr$ for three $1~\Gyr$ intervals \citep[see][]{Bahe+2017}. In this work, we only make use of full-size snapshots.
 
 The friends-of-friends groups and self-bound subhaloes in each snapshot are identified with SUBFIND \citep{Springel+2001,Dolag+2009}. To reconstruct the evolution of individual galaxies, subhaloes are linked in time with the \textsc{spiderweb} algorithm \citep{Bahe+2019}, which generates the merger tree of each individual galaxy.

\begin{figure*}
 \includegraphics[keepaspectratio, width=0.95\textwidth]{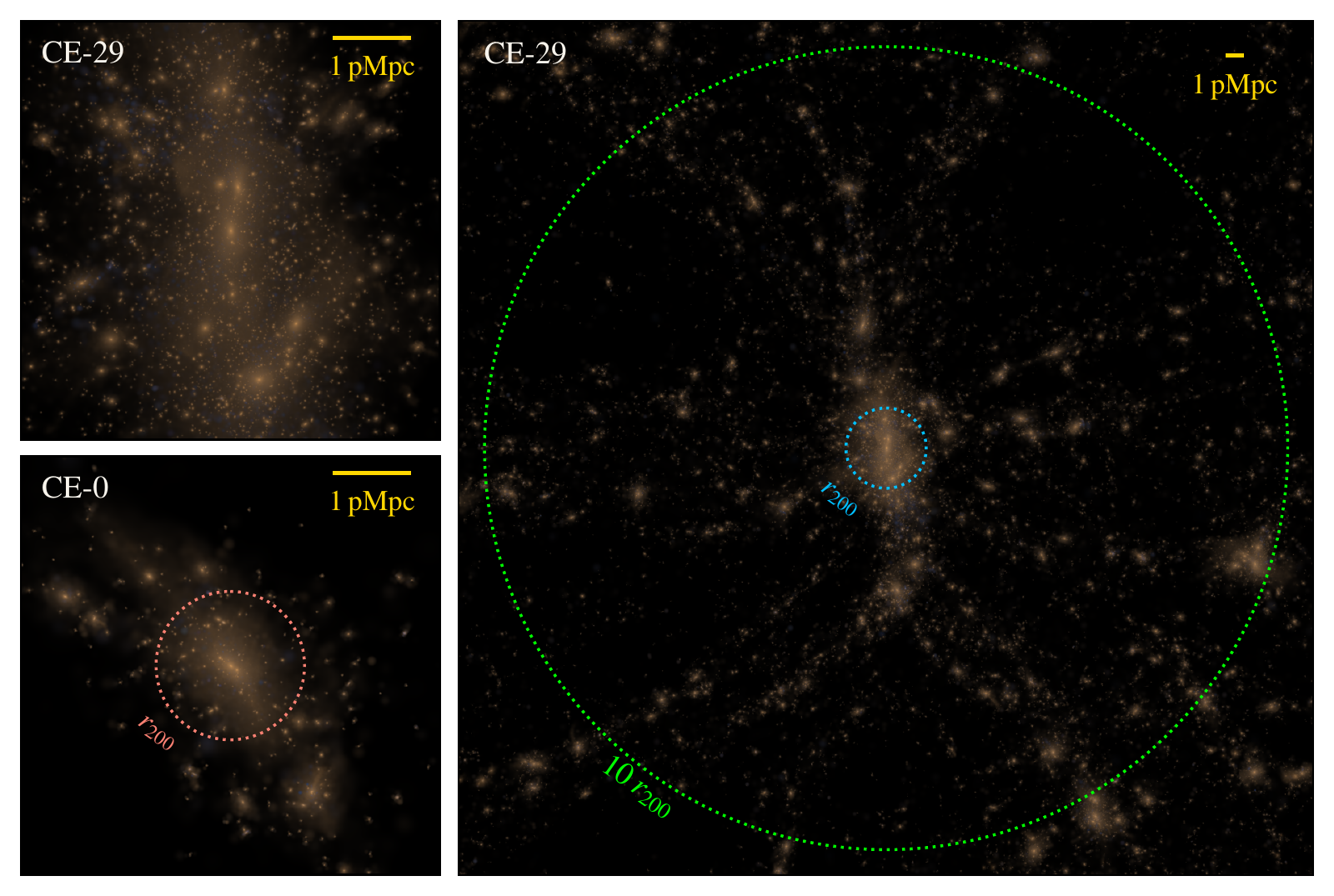}
 \caption{Composite \textit{gri} images of the stellar component for a low mass and the most massive C-EAGLE cluster at $z=0$, CE-0 ($M_{200} = 10^{14.07}~\Msun$ and $r_{200} = 1.03~\Mpc$) and CE-29 ($M_{200} = 10^{15.38}~\Msun$ and $r_{200} = 2.82~\Mpc$), respectively. Each image is a projection of all the stellar particles along the simulation $x$ axis, centred on the potential minimum of the central cluster galaxy. The right panel shows the entire high resolution volume for CE-29, marking the 1 and $10 \times r_{200}$ scales as blue and green dotted circles. The top left panel shows the region inside $r_{200}$ for CE-29, while CE-0 is shown with the same physical scale in the bottom left panel, along with its $r_{200}$ circle in red. The reference scale of $1~\Mpc$ is shown at the top right corner of each image as a yellow bar. In the image building process we employed the $C^2$ \citet{Wendland1995} kernel, where the smoothing length for each particle has been computed as the distance to its 58-th stellar neighbour. The three \textit{gri} images have been combined with the $\arcsinh$ scaling method of \citet{Lupton+2004}.}
 \label{fig_gri}
\end{figure*}

\section{C-EAGLE photometry and spectroscopy}\label{sec_photometry}

In this section, we detail the calculation of the spectrum and luminosity of star particles and galaxies in the simulations. The star particle's luminosity data are publicly available (see the Data Availability statement at the end of the paper).

\subsection{SED and luminosity computation}\label{sec_mag}
In this work, we limit ourselves to the modelling of the stellar emission, eventually dimmed by dust. We neglect dust re-emission in the far-infrared and light scattering on dust grains. This last approximation is supported by observations showing that scattering does not contribute significantly to the observed attenuation curve in galaxies \citep{Fischera+2003}, allowing us to treat each wavelength band independently. We post-processed the \CEAGLE simulation with a parallel, shared-memory python-C code developed for this analysis. The data has been computed in a series of 3D concentric spherical apertures, and by resampling recent star formation (see below for more details on the resampling).

We treated each stellar particle as a single stellar population (SSP), characterised by its initial mass, metallicity and age, stored at formation time in the simulation. In our analysis, we used the SPH-smoothed metallicity as metallicity \citep{Wiersma+2009}, which is the metallicity used by the model to define the density threshold for star formation. The average mass of the newborn star particles is $\sim 10^{6}~\Msun$. This resolution is sufficient to resolve in mass the SFRs typical of high-redshift, massive galaxies ($\sim 100~\Msunyr$), where plenty of cold gas is converted into stars in a relatively short time. On the other hand, it poorly resolves in time the stellar population generated at low-redshift or in low-mass galaxies, whose SFR is typically $< 1$--$10~\Msunyr$, i.e. the young stars are sampled too coarsely in time. 
In order to mitigate the stochasticity of the SF model, we resampled recent star formation following the method of \cite{Trayford+2015}. Every star particle with an age $< 100~\Myr$ is converted back into its parent gas particle, the SFR is recomputed, and by assuming it constant over the $100~\Myr$ interval, the final stellar mass produced is obtained. The latter is then resampled into a number of star particles of mass $10^4~\Msun$, with an age randomly chosen in the $6.3-100~\Myr$ interval, retaining the same metallicity of the original stellar particle. Finally, the resulting $10^4~\Msun$ stellar particles are included in the flux computation. We do the same with the star forming gas particles \citep{Schaye.DallaVecchia2008,Trayford+2015}\footnote{Note that this method differs from the modelling of \citet{Trayford+2017}.}.

We employed the state-of-the-art stellar population model E-MILES, based on a fully empirical stellar spectra library \citep{Vazdekis+2016} and the Padova isochrones \citep{Girardi+2000}. The E-MILES spectra extend from the far UV to the mid-IR ($1680~\text{\AA}$ to 5~\micron), with a constant $\Delta\lambda=0.9~\text{\AA}$, for a grid of stellar ages ranging from $6.3~\Myr$ to $17.8~\Gyr$, and metal mass fraction ranging $(4\times 10^{-4}-0.05)$ for SSPs younger than $63~\Myr$, and $(10^{-4}-0.03)$ otherwise. Every SSP in E-MILES is fully determined by specifying its age, metallicity, initial mass and initial mass function (IMF). We adopted the \citet{Chabrier2003} IMF, in accordance with the \CEAGLE star formation model, and we performed a bi-linear interpolation in age and metallicity of the E-MILES SEDs. Extrapolating the spectra outside the population synthesis model ranges can produce SEDs with negative fluxes, so we avoided any extrapolation by simply limiting the stellar particle's metallicities and ages.

For each simulated galaxy, we compute the full SED of each star particle, sum them within the selected spherical aperture, and apply dust attenuation to the integrated spectra. Finally, the SED is convolved with the response curves of the selected filters to obtain the galaxy broad-band luminosities.\footnote{In the case of the stellar particles luminosities, we compute only the ``un-obscured" broad-band luminosities.} %es by immediately convolving the particle SED with the filters' response  curves.}
%This allows us to compute both galaxy SEDs and magnitudes at the same time.

\subsection{Photometric systems}\label{sec_photometricSystem}

%We convolved each spectrum (either of an individual stellar particle or an entire galaxy) with a set of filter response curves, and integrated it to compute the corresponding luminositiy.
We adopted the following photometric systems: Sloan Digital Sky Survey (SDSS) \textit{ugriz} \citep{Fukugita+1996} with 1.2 air mass, Johnson-Cousins \textit{UBVRI},\footnote{\url{http://www.tng.iac.es/instruments/filters}} the UK Infrared Digital Sky Survey (UKIDSS) \textit{JHKYZ} \citep{Hewett+2006} with 1.3 air mass, 2MASS \textit{JHK$_s$} \citep{Cohen+2003} and JWST NIRCam\footnote{\url{https://jwst-docs.stsci.edu/near-infrared-camera/nircam-instrumentation/nircam-filters}} \citep{Beichman+2012}. The response curves have been downloaded from the Spanish Virtual Observatory website\footnote{\url{http://svo2.cab.inta-csic.es/theory/fps3}}, and later interpolated on the E-MILES wavelength grid. The zero-point luminosities were computed using the interpolated filter curves.

From each spectrum, we computed both the rest-frame and the observer's frame Pogson absolute AB magnitude, the latter by redshifting the spectrum according to the cosmological redshift of the snapshot, and maintaining the filters at rest. The E-MILES spectra are in units of $\Lsun\,\Msun^{-1}\,\text{\AA}^{-1}$, and the absolute magnitude in a generic band, $b$, is obtained as:
\begin{equation}
%M_\mathrm{b}^\mathrm{AB} = -2.5 \log _{10} \dfrac{L_\mathrm{b}^\mathrm{AB}}{L_\mathrm{b,0}^\mathrm{AB}}.
M_b^\mathrm{AB} = -2.5 \log _{10} \dfrac{L_b^\mathrm{AB}}{L_\mathrm{b,0}^\mathrm{AB}}.
\end{equation}
The luminosity in $b$ of a stellar particle of initial mass $m$ is computed as
\begin{equation}
%L_\mathrm{b}^\mathrm{AB} = m \int Q_\mathrm{b} F d\lambda,
L_b^\mathrm{AB} = m \int Q_b F d\lambda,
\end{equation}
%where $Q_\mathrm{b}$ is the filter response curve and $F$ is the spectrum interpolated from the EMILES library. The zero-point luminosity of the filter, $L_\mathrm{b,0}^\mathrm{AB}$, is obtained as
where $Q_b$ is the filter response curve and $F$ is the spectrum interpolated from the EMILES library (see the previous section). The zero-point luminosity of the filter, $L_{b,0}^\mathrm{AB}$, is obtained as 
\begin{equation}
%L_\mathrm{b,0}^\mathrm{AB} = 4\upi c D^2 \int Q_\mathrm{b} F_\mathrm{AB}  \dfrac{d\lambda}{\lambda^{2}},
L_{b,0}^\mathrm{AB} = 4\upi c D^2 \int Q_b F_\mathrm{AB}  \dfrac{d\lambda}{\lambda^{2}},
\end{equation}
where $c$ is the speed of light, $D=10~\pc$, and $F_\mathrm{AB} = 3631\times 10^{-23}~\mathrm{erg\,s^{-1}\,cm^{-2}\,Hz^{-1}}$ is the standard AB flux. All the filters considered here overlap with the rest-frame E-MILES range. We show in figure~\ref{fig_gri} composite \textit{gri} images of the stellar component for a light (CE-0) and massive (CE-29) cluster at $z=0$. The largest panel is a projection of all the stellar particles inside $10\times r_{200}$ of CE-29, while the left panels are close ups of the virial regions of both clusters with the same spatial scale.

\subsection{Dust attenuation}
Following the ISM dependant dust (GD) model of \citet{Trayford+2015}, we compute the intrinsic dust attenuation due to the ISM, but normalising the extinction law with the surface density of heavy elements in the gas phase, instead of their total mass. Thus, we multiply equation~(4) of \citet{Trayford+2015},
\begin{equation}
\begin{aligned}
\hat{\tau}_\mathrm{bc} &\rightarrow \dfrac{Z_\mathrm{SF}}{Z_{Z14}(\Mstar = M_\mathrm{MW})} \hat{\tau}_\mathrm{bc} \\[2ex]
\hat{\tau}_\mathrm{ism} &\rightarrow \dfrac{Z_\mathrm{SF}}{Z_{Z14}(\Mstar = M_\mathrm{MW})} \hat{\tau}_\mathrm{ism},
\end{aligned}
\label{eq_dust}
\end{equation}
by $(M_\mathrm{SF}/M_\mathrm{MW})\times(R_\mathrm{MW}/R_\star)^2$,
%
%\begin{equation}
% \dfrac{M_\mathrm{SF}}{M_\mathrm{MW}} \left( \dfrac{R_\mathrm{MW}}{R_\star} \right)^2, \label{eq_dust}
%\end{equation}
%
where $M_\mathrm{SF}$ is the mass of star forming gas contained in the galaxy half-mass stellar radius $R_\star$, $M_\mathrm{MW}=8.43\times 10^9~\Msun$ is half of the cold ISM mass in the Milky Way, and $R_\mathrm{MW}=5~\kpc$ is the Milky Way half-light radius \citep{Li2016}. Thus, our dust attenuation law can increase (decrease) the galaxy luminosity, with respect to \citet{Trayford+2015}, if the cold gas average surface density is low (high). We computed the $r$ magnitude difference of the original GD model of \citet{Trayford+2015} and our dust prescription for all the \CEAGLE galaxies bound to the main haloes at $z=0$, having $r<r_{200}$ and $\Mstar > 10^8~\Msun$, resulting in 25847 galaxies, with only 904 having $\lvert M_r - M_r^\mathrm{T} \rvert > 0.01~\mathrm{mag}$. This effect is maximal for central galaxies. These are up to 1.5~mag brighter with our dust attenuation formulation, despite the fact that they contain a significant amount of cold gas, because their stellar half mass radius is large. On the other hand, for $\Mstar < 10^9~\Msun$, the higher gas densities produce few tens of less luminous galaxies, with a maximum dimming of 2~mags. Galaxies with $\Mstar \sim 10^{11}~\Msun$ show only a slight deviation from \citet{Trayford+2015}. The $g-r$ colour of central galaxies in the C-EAGLE sample ($\Mstar \gtrsim 10^{12}~\Msun$) is affected as well. The GD model produces central galaxies that are much redder than the red sequence, up to $g-r = 1.1$, with outliers in the $2-18 \sigma$ range. Our model yields central galaxies colours that are in accordance with the red sequence.

All the spectra are computed within a spherical three-dimensional aperture of 1, 3, 5, 10, 20, 30, 40, 50, 70 and $100~\mathrm{pkpc}$. In addition, we computed the stellar velocity dispersion inside each 3D aperture, and inside the galaxy projected half-mass radius along each axis. We calculated the integrated spectra for each galaxy identified by the SUBFIND algorithm having a total stellar mass larger than $10^{7.5}~\Msun$, in order to remove the vast majority of subhaloes resolved by less than few tens of particles.

\section{Building mock luminosity functions}\label{sec_3}
\subsection{Galaxies selection}\label{galSelection}

The galaxy stellar luminosity function essentially counts how many objects fall in a bin of magnitude per unit area in the sky (mainly in case of observations) or volume (most used in theoretical works). In the case of poorly populated clusters, their LF will be noisy. Stacking several clusters helps both increasing the galaxy counts and enhancing common trends among different systems. In order to compare the simulated clusters with observations, we stack the \CEAGLE clusters together (see section~\ref{sec_observations}). We adopted the magnitudes calculated in a 3D spherical aperture of $30~\mathrm{pkpc}$, and we include in the magnitude computation only the stellar particles that are gravitationally bound to the galaxy. As shown by \citet{Schaye+2015}, this yields stellar masses comparable to those inferred accounting the mass within a projected circular aperture of the Petrosian radius.

\begin{figure}
\centering
\includegraphics[keepaspectratio, width=0.4\textwidth, trim=25 0 0 0]{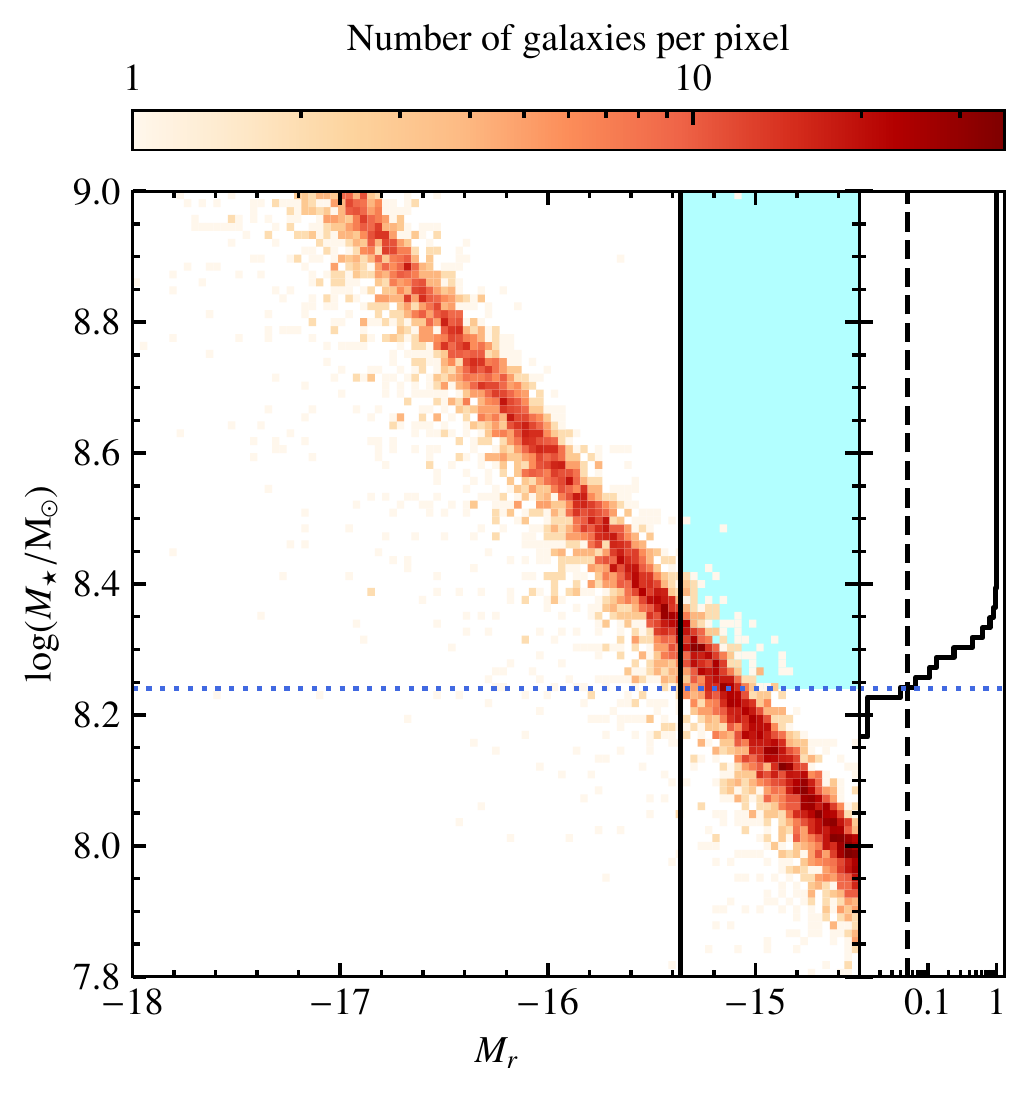}
 \caption{Left panel: magnitude-stellar mass relation for all the galaxies in \CEAGLE at $z=0$, only the low-mass end is shown. The blue dotted line shows our stellar mass cut at $\Mstar \simeq 1.8\times 10^8~\Msun$, and the solid vertical line is the corresponding magnitude (see section~\ref{galSelection}). The light blue area is the region above the mass threshold where the magnitude-stellar mass relation is incomplete. Right panel: normalized cumulative distribution of galaxy masses along the solid vertical line, the dashed line marks the distribution's 5th percentile.}
 \label{mass_mag}
\end{figure}
The finite resolution of the \CEAGLE simulation implies that low-mass subhaloes are resolved by a small number of particles, therefore prompting for the use of a stellar mass cut, which, in turn, introduces incompleteness in a magnitude-selected sample. We show in figure~\ref{mass_mag} the stellar mass-magnitude relation for all the \CEAGLE galaxies inside $r_{200}$ at $z=0$ having $7.8<\log (\Mstar/\Msun) < 9$, where we adopt as a mass limit of 100 times the gas particles initial mass, corresponding to $\Mstar \simeq 1.8 \times 10^8~\Msun$ (dotted red line in the figure). The cut in magnitude should not be in the region of the plot where the mass-magnitude relation is incomplete due to the mass cut (blue region in the figure). The magnitude limit is determined as follows: we bin the galaxies into 200 bins in magnitude in the $(-18,-14.5)$ range, and we compute the cumulative mass distribution for each bin. The magnitude limit is defined as the magnitude of the faintest bin having its mass distribution 5th percentile above $\simeq 1.8 \times 10^8~\Msun$. The right panel of figure~\ref{mass_mag} shows the cumulative distribution in mass of the bin centered on $-15.36$ as a solid line, with a vertical dashed line marking the 5th percentile. This procedure is perfomed by stacking all the clusters at every snapshot.

In the computation of the LF, we remove from the sample the brightest galaxy of each cluster, as commonly done in observations \citep{Hansen+2005,Wen.Han2015}. We also prevent contamination from low-resolution particles (massive dark matter particles residing outside the high-resolution region, in the initial conditions) by discarding all galaxies that are closer than the minimum of $[1~\mathrm{pMpc}, r_{200}]$ to such low-resolution particles, where $r_{200}$ is the virial radius of the friends-of-friends group to which the galaxies belong.

\subsection{Luminosity functions computation}\label{sec_lumFunc}
Historically, LFs have been fit with a single \citet{Schechter1976} function, until recent deeper observations have shown a steepening of the faint end \citep{Blanton+2005,Baldry+2008,Li.White2009,Drory+2009,Moustakas+2013,Muzzin+2013, Tomczak+2014}, which requires a double Schechter function or a combination of Schechter function and a power-law to be fit. A fit with two Schechter functions, having six free parameters, can be highly degenerate. We follow the approach already used in previous work \citep{Blanton+2005,Mortlock+2015,Agulli+2016,Annunziatella+2016} thus employ two Schechter functions with the same knee magnitude, $M^*$, lowering the number of variables to 5. Often, in observations LFs are fit with a single Schechter, especially when the limiting absolute magnitude does not reach well beyond -18~mag. For the sake of completeness, and to ease the comparison of synthetic LFs with observations, we fit every LF with both a single and double Schechter functions, respectively given by:
\begin{equation}
 \Phi_1(M) = 0.4 \ln(10) \phi 10^{0.4(\alpha +1) (M^* - M)} \exp\left[-10^{0.4(M^*-M)} \right],
\end{equation}
where $M^*$, $\alpha$ and $\phi$ are the knee magnitude, faint-end slope and normalization of the single Schechter function, $\Phi_1$, and
\begin{multline}
 \Phi_2(M) = 0.4 \ln(10)\exp\left[-10^{0.4(M_2^*-M)} \right] \\[2ex]
 \left[ \phib 10^{0.4(\alphab +1) (M_2^* - M)} + \phif 10^{0.4(\alphaf +1) (M_2^* - M)} \right],
\end{multline}
where $\alphab$, $\phib$, $\alphaf$ and $\phif$ are the slope and normalization of the bright and faint end of the double Schechter function, $\Phi_2$, and $M_2^*$ is the mutual knee magnitude.

Determining the LF from a set of magnitudes means to recover the galaxy distribution in magnitude. Usually, the data is binned in magnitude, and to each bin an approximate Poisson error is assigned. %\footnote{$\sqrt{n}$ where $n$ is the number of galaxies in a bin.}.
The resulting histogram is then fit with a function (in our case $\Phi_1$ or $\Phi_2$). \citet{Andreon+2005} show that binning data in such way, although being very useful for a graphical representation, can lead to biases and a loss of information. In particular, the number of datapoints decreases considerably when binning is adopted, e.g. from 3288 to just 16 in the $r$ band LF in figure~\ref{fig_LF1}. The Bayesian method exemplified here (or even a maximum likelihood method) employs instead all the data, having the potential of reducing the uncertainties of the inferred parameters. Therefore, we employed Bayesian inference to compute the cluster LFs. In particular, we adapted the approach described in \citet{Andreon+2005}, based on the extended likelihood, which accounts also for the sample selection bias produced by the cut in magnitude. The extended likelihood is defined as follows:
\begin{equation}
 \ln \Lambda = \sum_{j=1}^C \left[ s_j + \sum_{i=1}^{N_j} \ln p(M_i) \right], \qquad s = -\int^{M_\mathrm{f}}_{M_\mathrm{b}} p(M) dM
\end{equation}
where $C$ is the number of clusters in the sample, $N_j$ is the total number of galaxies of cluster $j$ being fitted, $M_i$ is the magnitude of the \textit{i}th galaxy, $p(M_i)$ is the extended (since its integral is not 1) probability of the \textit{i}th galaxy of the \textit{j}th cluster to have magnitude $M_i$, and $p(M)$ is either equal to $V_j\Phi_1$ or $V_j\Phi_2$, where $V_j$ is the cluster volume (in this case the sphere or spherical shell volume). In this work we clearly do not suffer from background contamination, since we are able to select the galaxies inside the simulated cluster, so we do not consider the additional term in the likelihood accounting for background sources, as in \citet{Andreon+2005}. The quantity $s$ is the expected number of galaxies given the model, where $M_\mathrm{f}$ is the faint-end limiting magnitude of the sample (in our study, constant for all the clusters, at a given redshift). $M_\mathrm{b}$ is the limiting magnitude at the bright end, which theoretically can be equal to $-\infty$ since our sample is not bounded in magnitude at the bright end. We assumed $M_\mathrm{b} = -100$ to ease the numerical convergence of $s$.

We determine the shape of the probability distribution of the model parameters with the Markov Chain Monte Carlo (MCMC) method using an affine-invariant ensemble sampler as implemented in the Python package \textsc{emcee} \citep{Foreman-Mackey+2013}. We adopted flat priors for the parameters, namely $-2.5<\alpha < 20$, $-26<M^*<-18$, $0<\phi<10^5$ for $\Phi_1$, and $-26<M_2^*<-18$, $-2.5<\alphaf < \alphab <20$, $0<\phif<\phib<10^5$ for $\Phi_2$, which ensure that the LF's bright-end is described by $\alphab$ and $\phib$. In order to describe the posterior distribution and give the best fit parameters, we report in the tables the 16th, 50th and 84th percentiles of the marginalized posterior for each fit.

As we proceed in our analysis, we need to select between the single- and double-Schechter fitting functions depending on which better represents the data, and taking into account that a larger number of free parameters can result in overfitting the data.
%In order to compares different functional fits, one  has to take into account the their number of free parameters, penalising complexity to avoid overfitting.
%A proper comparison is often a complicated task, and the data can range from clearly favouring a the simpler or the most complex one.
We adopt here the Bayesian Information Criterion \citep[hereafter BIC]{Schwarz1978,Gregory2005} to select between a single- and a double-Schechter fitting function. The goodness of the fit is described by the BIC index, which is formally defined as $k\ln (n) - 2\ln (\Lambda_\mathrm{max})$, where $k$ is the number of parameters in the model, $n$ is the number of data points, and $\Lambda_\mathrm{max}$ is the maximized value of the likelihood. The factor $k\ln(n)$ modulates the total index with increasing complexity. Different ranges of $\Delta \mathrm{BIC}$ indicate a positive (2--6), strong (6--10) and very strong (larger than 10) evidence against the fitted model \citep{Gregory2005}.

It is important to note that, while the BIC analysis might favour a double Schechter function in some cases, the value of the faint-end slope can be still similar to that of the LF of field galaxies. In other words, requiring a double Schechter does not automatically imply a substantially steeper faint end. However, we have found that for each pair of fits on the same data, the double Schechter function always presents a brighter magnitude at the knee and a steeper faint-end, so that, forcing the fit to be a single Schechter on double-slope data, can lead to underestimating the faint-end slope.

\subsection{Code validation: recovering the \EAGLE luminosity function}\label{sec_code_validation}
In order to validate both the post-processing method and Bayesian derivation of the luminosity functions, we apply our analysis to the \EAGLE simulation and compare it with the data provided by the \EAGLE database.
\begin{figure}
\centering
 \includegraphics[keepaspectratio, width=0.4\textwidth, trim=20 0 0 0]{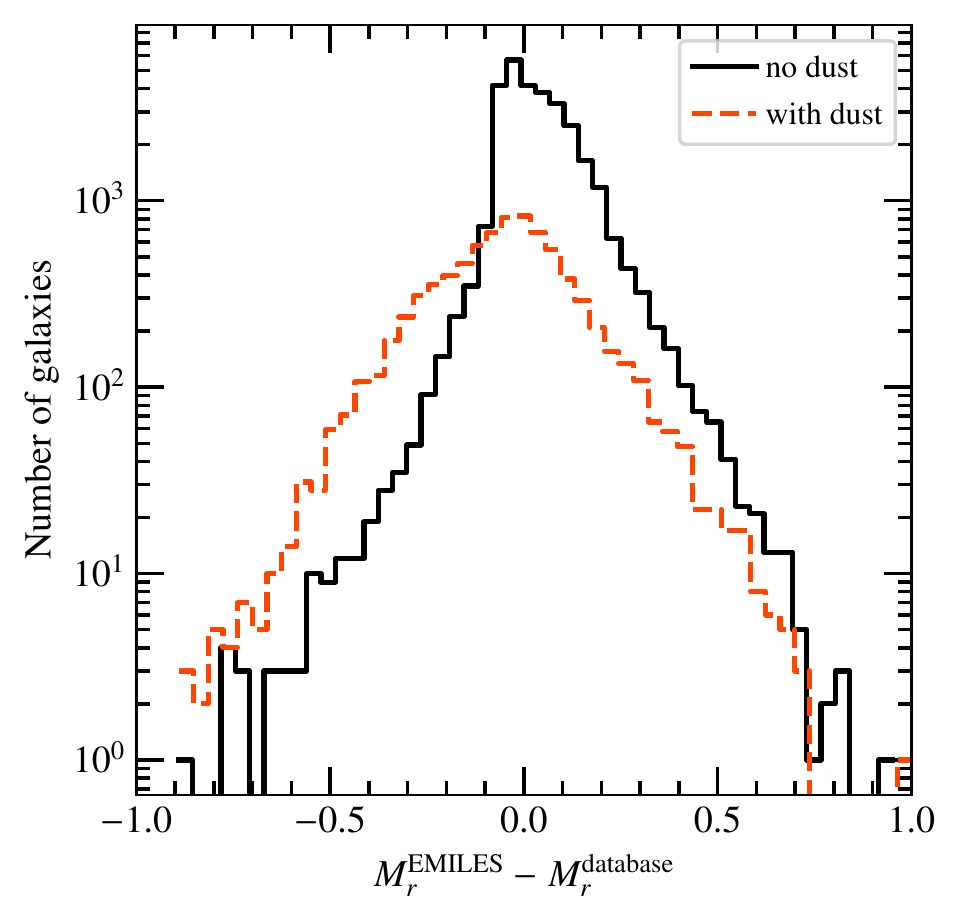}
 \caption{Comparison between the Ref \EAGLE simulation data at $z=0.1$ and the post-processing based on the E-MILES spectral library. The distribution of matching galaxies' SDSS $r$-band magnitudes difference between the two models is plotted. Solid and dashed lines are for the dust unobscured and obscured samples, respectively.}
 \label{fig_EAGLEcomparison}
\end{figure}
\begin{table}
\centering
\caption{Best single Schechter fit parameters for the Ref \EAGLE simulation in the $r$ band at $z=0.1$. The columns list the limiting magnitude $M_\mathrm{f}$, knee magnitude $M_r^*$, faint end slope $\alpha$, and normalization $\phi$. The full table is available in the online material.}\label{tabEAGLE}
\begin{tabular}{ccccccccccc}
\toprule
$M_\mathrm{f}$& $M_r^*$                & $\alpha$                  & $1000\times\phi$\\
      (mag)        &  (mag)                 &                           & ($h^3~\cMpc^{-3}~\mathrm{mag}^{-1}$)\\
% (1) & (2) & (3) & (4) &(5) \\
\midrule

 $-19.0$ & $-21.28^{+0.05}_{-0.05}$ & $-1.11^{+0.03}_{-0.03}$ & $11.89^{+0.76}_{-0.73}$\\[1.3ex]
 $-18.5$ & $-21.32^{+0.05}_{-0.05}$ & $-1.14^{+0.02}_{-0.02}$ & $11.33^{+0.62}_{-0.61}$\\[1.3ex]
 $-18.0$ & $-21.35^{+0.04}_{-0.04}$ & $-1.16^{+0.02}_{-0.02}$ & $10.85^{+0.51}_{-0.50}$\\[1.3ex]
 $-17.5$ & $-21.40^{+0.04}_{-0.04}$ & $-1.19^{+0.01}_{-0.01}$ & $10.12^{+0.42}_{-0.40}$\\[1.3ex]
 $-17.0$ & $-21.46^{+0.04}_{-0.04}$ & $-1.22^{+0.01}_{-0.01}$ & $9.39 ^{+0.36}_{-0.35}$\\[1.3ex]
 $-16.5$ & $-21.54^{+0.04}_{-0.04}$ & $-1.25^{+0.01}_{-0.01}$ & $8.50 ^{+0.29}_{-0.29}$\\[1.3ex]
\bottomrule
\end{tabular}
\end{table}

\begin{table}
\centering
\caption{Same as table~\ref{tabEAGLE}, but at $z=0$. The full table is available in the online material.}\label{tabEAGLEcont}
\begin{tabular}{ccccccccccc}
\toprule
  $M_\mathrm{f}$& $M_r^*$                & $\alpha$                  & $1000\times\phi$\\
      (mag)        &  (mag)                 &                           & ($h^3~\cMpc^{-3}~\mathrm{mag}^{-1}$)\\
% (1) & (2) & (3) & (4) &(5) \\
\midrule
$-19.0$ & $-21.19^{+0.06}_{-0.06}$ & $-1.12^{+0.04}_{-0.04}$ & $11.02^{+0.78}_{-0.72}$\\[1.3ex]
$-18.5$ & $-21.21^{+0.05}_{-0.05}$ & $-1.14^{+0.03}_{-0.03}$ & $10.72^{+0.65}_{-0.60}$\\[1.3ex]
$-18.0$ & $-21.24^{+0.05}_{-0.04}$ & $-1.16^{+0.02}_{-0.02}$ & $10.30^{+0.51}_{-0.50}$\\[1.3ex]
$-17.5$ & $-21.28^{+0.04}_{-0.04}$ & $-1.18^{+0.01}_{-0.01}$ & $9.85 ^{+0.44}_{-0.43}$\\[1.3ex]
$-17.0$ & $-21.33^{+0.04}_{-0.04}$ & $-1.21^{+0.01}_{-0.01}$ & $9.21 ^{+0.36}_{-0.35}$\\[1.3ex]
$-16.5$ & $-21.42^{+0.04}_{-0.04}$ & $-1.25^{+0.01}_{-0.01}$ & $8.16 ^{+0.30}_{-0.29}$\\[1.3ex]
\bottomrule
\end{tabular}
\end{table}

\begin{figure}
\centering
 \includegraphics[keepaspectratio, width=0.45\textwidth, trim=10 0 0 0]{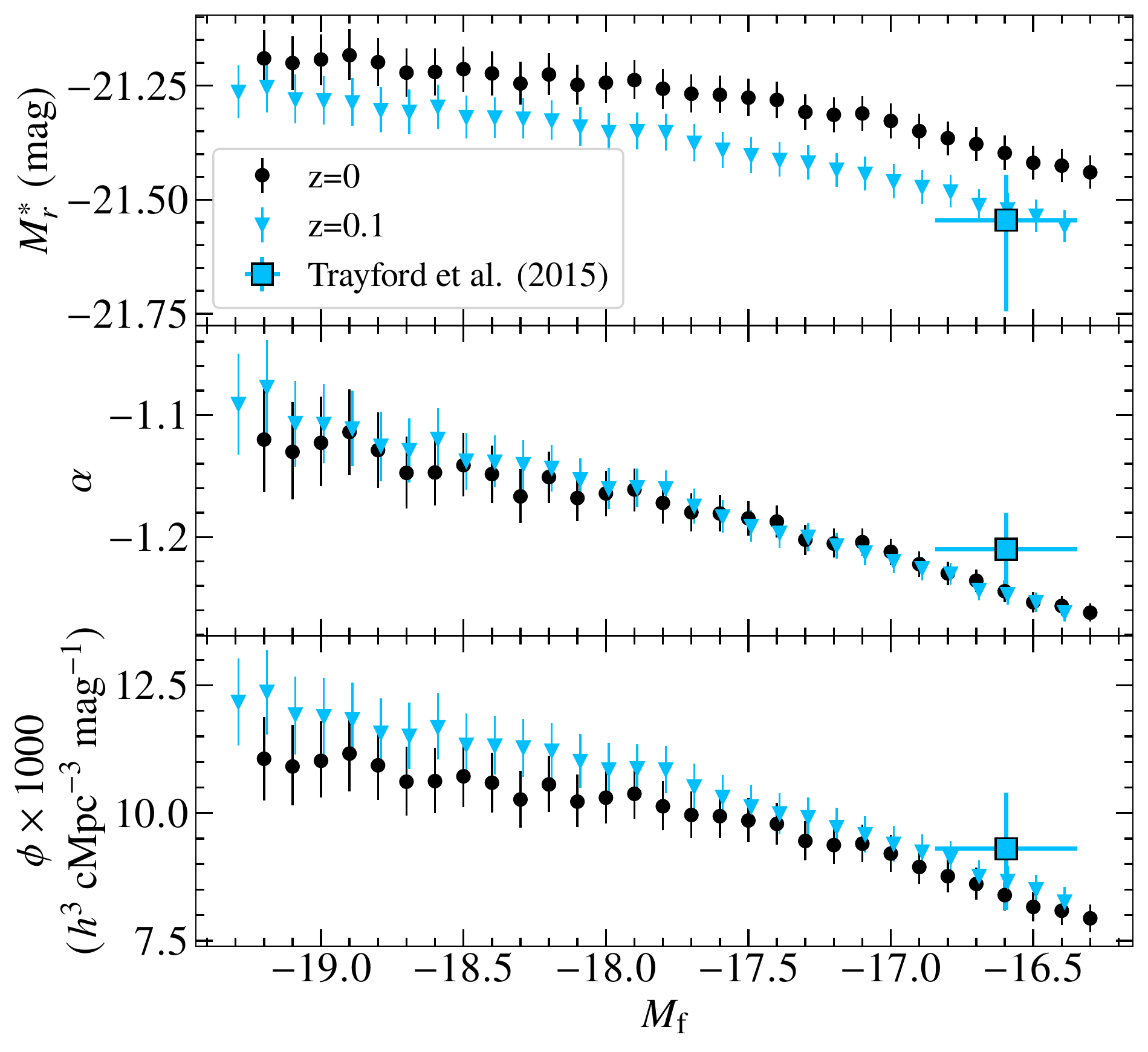}
 \caption{Best fit parameters for the \EAGLE simulation in the $r$-band at $z=0.1$ and 0 (blue triangles and black circles, respectively) as a function of the faint-end limiting magnitude $M_\mathrm{f}$. The data span from $M_\mathrm{f}$ corresponding to 100 to 1000 particles in the $30~\kpc$ aperture (see tables~\ref{tabEAGLE} and \ref{tabEAGLEcont}). The blue square is the result of \citet{Trayford+2015} at $z=0.1$, with the horizontal error bar representing the width of their magnitude bin.}
 \label{fig_EAGLEfits}
\end{figure}
We processed the Ref \EAGLE simulation snapshot at $z=0.1$, obtaining the absolute magnitudes of 78275 galaxies having stellar mass larger than $10^{7.5}~\Msun$. We show in figure~\ref{fig_EAGLEcomparison} the difference in the $r$-band absolute magnitude between our E-MILES post-processed galaxies and the exact same sample of galaxies taken from the \EAGLE database. We note that, in \EAGLE, dust attenuation is applied only to galaxies having more than 250 sub-particles of dust, resulting in a smaller number of objects. We cross match the galaxies with dust attenuation in the EAGLE database with our sample, to perform a fair comparison. The median of the distribution is $0.02~\mathrm{mag}$ $(-0.04~\mathrm{mag}$) without dust correction (with dust correction), whereas the standard deviation is $0.12~\mathrm{mag}$ ($0.19~\mathrm{mag}$). Therefore, the method well matches the magnitudes of the \EAGLE galaxies. The larger scatter when dust attenuation is applied is due to the different recipe employed (see section~\ref{sec_mag}), and the scatter does not depend strongly on the galaxy's magnitude, being smaller for faint objects due to the lack of cold, star forming gas.

We then tested the LF computation method by computing the LF of all galaxies in the sample. To be consistent with the literature, we focussed solely on the single-Schechter fit, which is appropriate in the case of field galaxies. Having fixed the galaxy sample, we need to select an appropriate limiting magnitude for the fit (see section~\ref{sec_lumFunc}). Caution in the selection of $M_\mathrm{f}$ is needed, since its choice can influence the faint-end slope, especially if the true underlying distribution is not a perfect Schechter function. We explored different values for the faint-end limiting magnitude $M_\mathrm{f}$: we selected a magnitude range corresponding to a stellar mass inside a $30~\kpc$ aperture of $1.8\times 10^8$ and $1.8\times 10^9~\Msun$ (100 and 1000 times the initial gas particles mass), computed with the same method shown in section~\ref{galSelection}. The results are summarised in tables~\ref{tabEAGLE} and \ref{tabEAGLEcont}, for different choices of limiting magnitude. We show in figure~\ref{fig_EAGLEfits} the fit parameters dependence on $M_\mathrm{f}$: a stronger magnitude cut produces the expected flattening of the faint end slope, together with a slightly reduced knee luminosity. The \citet{Trayford+2015} best fit at $z=0.1$ is the blue square in the figure, having $M_r^*= -21.57^{+0.036}_{-0.035}$, $\alpha = -1.21 \pm 0.03$, and $\phi = 9.3_{-1.1}^{+1.2} \times 10^{-3}h^3~\cMpc^{-3}~\mathrm{mag}^{-1}$. While our results matches \citet{Trayford+2015} in the case of $M_r^*$ well, the slope $\alpha$ is slightly steeper and $\phi$ is lower, although still within the errors. The parameters $\alpha$ and $\phi$ of \citet{Trayford+2015}'s analysis are best matched for $M_\mathrm{f}\simeq -17$, allowing for a difference of about $0.15~\mathrm{dex}$ in the knee magnitude between the two models.

\section{Results}\label{sec_results}
\begin{figure*}
 \includegraphics[keepaspectratio, width=0.95\textwidth]{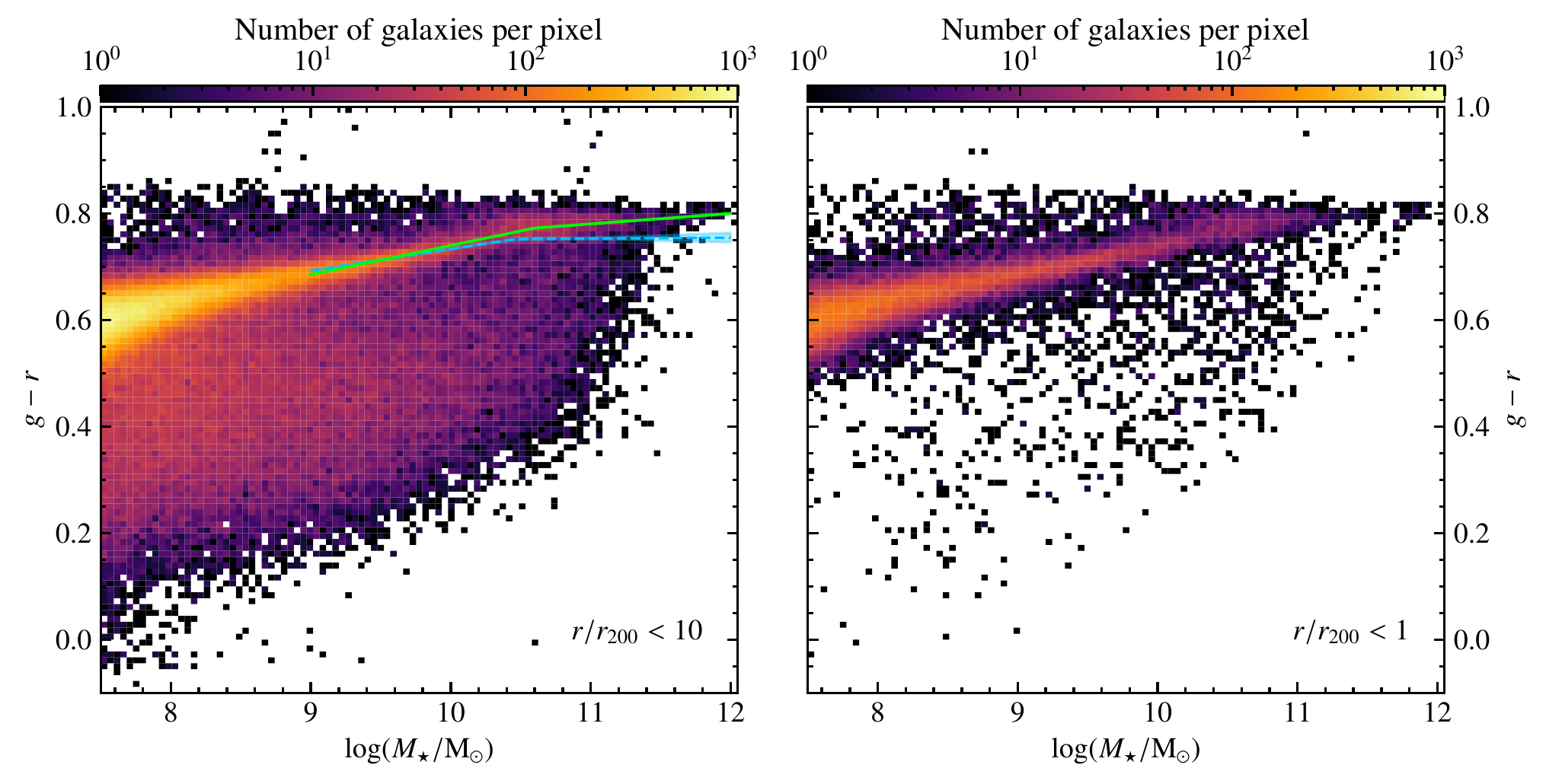}
 \caption{Rest-frame, dust attenuated $g-r$ colour-stellar mass diagram for the stacked 30 \CEAGLE simulations at $z=0.1$. Galaxies magnitudes and stellar masses are computed inside an aperture of $30~\kpc$. The left panel shows all the galaxies having $r/r_{200}<10$, in the case of the \textit{Hydrangea} sample, or $r/r_{200}<5$ for the remaining 6 clusters of \CEAGLE; the right panel shows only the virial regions of the cluster ($r/r_{200}<1$). The colour bar represents the galaxy number density in each bin of colour and stellar mass, and the solid green and dashed blue lines are the \CEAGLE and \EAGLE best fit of their respective red sequences (galaxies having $\Mstar<10^9~\Msun$ are excluded from the fit).}
 \label{color_mass}
\end{figure*}
In this section we present the results for the sample of 30 simulated galaxy clusters in \CEAGLE. We first show the galaxies' photometric properties, discussing their colour-stellar mass relation. We then present and discuss the LFs of the \CEAGLE sample at $z=0$. Finally, we compare the \CEAGLE LFs with observations of individual (section~\ref{sec_individual_clusters}) and stacked clusters (section~\ref{sec_stacked_clusters}).

\subsection{Colour-stellar mass relation}\label{sec_colormass}
\begin{table}
\centering
\caption{Best fit parameters for the \EAGLE and \CEAGLE red sequence fitting. A machine readable version is available in the online material.}\label{tabRedSeqFit}
\begin{tabular}{ccccccccccc}
\toprule
parameter & \EAGLE & \CEAGLE \\
\midrule

$a_1$ & $40.71_{-1.19}^{+1.28}\times 10^{-3}$        &$54.66_{-0.430}^{+0.430} \times 10^{-3}$\\[1.3ex]
$b$   & $32.63_{-1.21}^{+1.13} \times 10^{-2}$       &$19.34_{-0.410}^{+0.407} \times 10^{-2}$\\[1.3ex]
$a_2$ & $1.73_{-6.97}^{+7.28}\times 10^{-3}$         &$19.57_{-1.57}^{+1.44}\times 10^{-3}$\\[1.3ex]
$m_0$ & $10.46_{-0.11}^{+0.07}$                      &$10.60_{-0.023}^{+0.026}$\\[1.3ex]
$\sigma_1^2$ & $17.18_{-1.02}^{+1.09}\times 10^{-5}$   &$13.54_{-0.33}^{+0.33}\times 10^{-5}$\\[1.3ex]
$\sigma_2^2$ & $33.08_{-1.16}^{+1.25}\times 10^{-4}$ &$34.75_{-0.64}^{+0.65}\times 10^{-4}$\\[1.3ex]
$\mu_2$ & $-33.56_{-1.64}^{+1.59}\times 10^{-3}$     &$-18.43_{-0.85}^{+0.81} \times 10^{-3}$\\[1.3ex]
\bottomrule
\end{tabular}
\end{table}
In figure~\ref{color_mass} we show the rest frame, dust attenuated $g-r$ colour-stellar mass diagram for all the 30 \CEAGLE clusters, down to a $30~\kpc$ aperture stellar mass of $10^{7.5}~\Msun$. In the left panel we plot all the galaxies inside the high resolution volume of each simulation, i.e. having $r/r_{200} < 10$ for the \textit{Hydrangea} sample and $r/r_{200} < 5$ for the additional 6 clusters, where $r$ is the distance from the centre of the most massive halo. In the right panel, we consider only the galaxies within the virial radius. The colour bars indicate the galaxy number in each $g-r$ colour and mass bin. We note that a blue cloud is visible when all galaxies are selected (left panel). It extends up to $10^{10.8}~\Msun$ in mass, with $g-r\simeq 0.15$ at $10^{9}~\Msun$. The blue cloud is absent when only galaxies within the virial radius are considered (right panel), in agreement with the high fractions of quenched galaxies found by \citet{Bahe+2017}. 

Both samples displays a tight red sequence across the entire mass range, that flattens at high masses. In order to compare it with that in \EAGLE, we fit the red sequence with a broken power-law for both \EAGLE and \CEAGLE. We consider only quiescent galaxies, defined as having a specific star formation rate $\mathrm{sSFR}<10^{-11}~\yr^{-1}$ at $z=0.1$ and, in the case of \CEAGLE, residing inside a radius of $10~r_{200}$ from the cluster centre. Furthermore, we exclude galaxies with $\Mstar<10^9~\Msun$, due to the unrealistic overabundance of red dwarfs \citep{Trayford+2015}. We adopt the broken power-law function:
\begin{equation}
f(x) = 
\begin{cases}
& a_1 x + b, \qquad  x \leq m_0, \\
& a_2 x + (a_1-a_2)m_0 + b, \qquad  x>m_0, \\
\end{cases}
\end{equation}
where $m_0$ is the logarithm of the stellar mass where the broken power-law changes slope. Selecting only quiescent galaxies removes the vast majority of the blue cloud, although some galaxies with colour in the range $0.4 < g-r < 0.6$ remain. Instead of artificially removing them, we take those galaxies into account by assuming that the generative function of our data is composed by the sum of a narrow Gaussian having a median equal to 0, modelling the dispersion of the red sequence, and a broader Gaussian that describes the bluer outliers. We later marginalize over the components of the second Gaussian. Thus, the log-likelihood $\Lambda_\mathrm{G}$ is
\begin{equation}
\ln \Lambda_\mathrm{G} = \sum \ln \left[ \dfrac{1}{2\sigma_1\sqrt{2\upi}}e^{-\frac{u^2}{\sigma_1^2}} + \dfrac{1}{2\sigma_2\sqrt{2\upi}} e^{-\frac{(u-\mu_2)^2}{\sigma_2^2}} \right],
\end{equation}
where $u=(g-r) - f(\log(\Mstar))$ is the data residuals given the model $f(x)$ and a certain choice of the parameters, $\sigma_1$ is the standard deviation of the red sequence, and $\mu_2$ and $\sigma_2$ are the mean and standard deviation of the broader Gaussian. As a prior we adopted $-1000<b<1000$, $0<\sigma_2<\sigma_1$, $-10<\mu_2<10$, $0<\arctan(a_2)< \arctan(a_1)< \upi/2$, $10<m_0<11$. In table~\ref{tabRedSeqFit}, we give the parameters of the best fit, together with the 16th, 50th and 84th percentiles of the marginalized posterior for each fit (the plots of the posteriors are available in the supplementary data). In figure~\ref{color_mass}, we plot the red sequence fit for \CEAGLE (green solid line) and \EAGLE (dashed blue line), where the shaded region represent the uncertainty. 
The red sequence does not have a constant colour, but becomes redder with increasing stellar mass up to $\log(\Mstar/\Msun) = 10.60^{+0.023}_{-0.026}$, reaching $g-r \gtrsim  0.75$ at  $\Mstar = 10^{10.5}~\Msun$, being $0.05~\mathrm{dex}$ redder than \EAGLE at $10^{11.5}~\Msun$. The most massive galaxies, typically centrals, have a $g-r \simeq 0.8$.

The virial region (right panel of figure~\ref{color_mass}) shows a well defined red sequence, and the absence of the blue cloud. As expected the majority of massive ($\Mstar>10^{10.2}~\Msun$) red galaxies reside in the virial region of the cluster, while the blue cloud in the left panel is dominated by galaxies in the clusters outskirts. Nevertheless, a large number of blue, star forming galaxies with $\Mstar>10^{9}\Msun$ are present in the virial regions and, in the same fashion as \EAGLE, \CEAGLE show a certain amount of massive blue galaxies ($\Mstar > 10^{11}~\Msun$, $0.5<g-r<0.7$), a minority of them inside the cluster's virial radius. Finally, all the 30 central galaxies are concentrated at the very massive end of the red sequence ($\Mstar \simeq 10^{12}~\Msun$), with $ 0.75<g-r<0.82$.

The reason why the red sequence of \CEAGLE galaxies is redder is twofold. All (massive) galaxies in \CEAGLE form and evolve in a higher-density environment by construction. The \EAGLE volume, because of its limited size, cannot contain such larger overdensities. Moreover, the efficiency of the AGN feedback has been increased in \CEAGLE, with respect to the original \EAGLE model, yielding faster quenching (and redder colour) of massive galaxies.

\subsection{C-EAGLE luminosity functions at z=0}
\begin{figure*}
\centering
 \includegraphics[keepaspectratio, width=\textwidth]{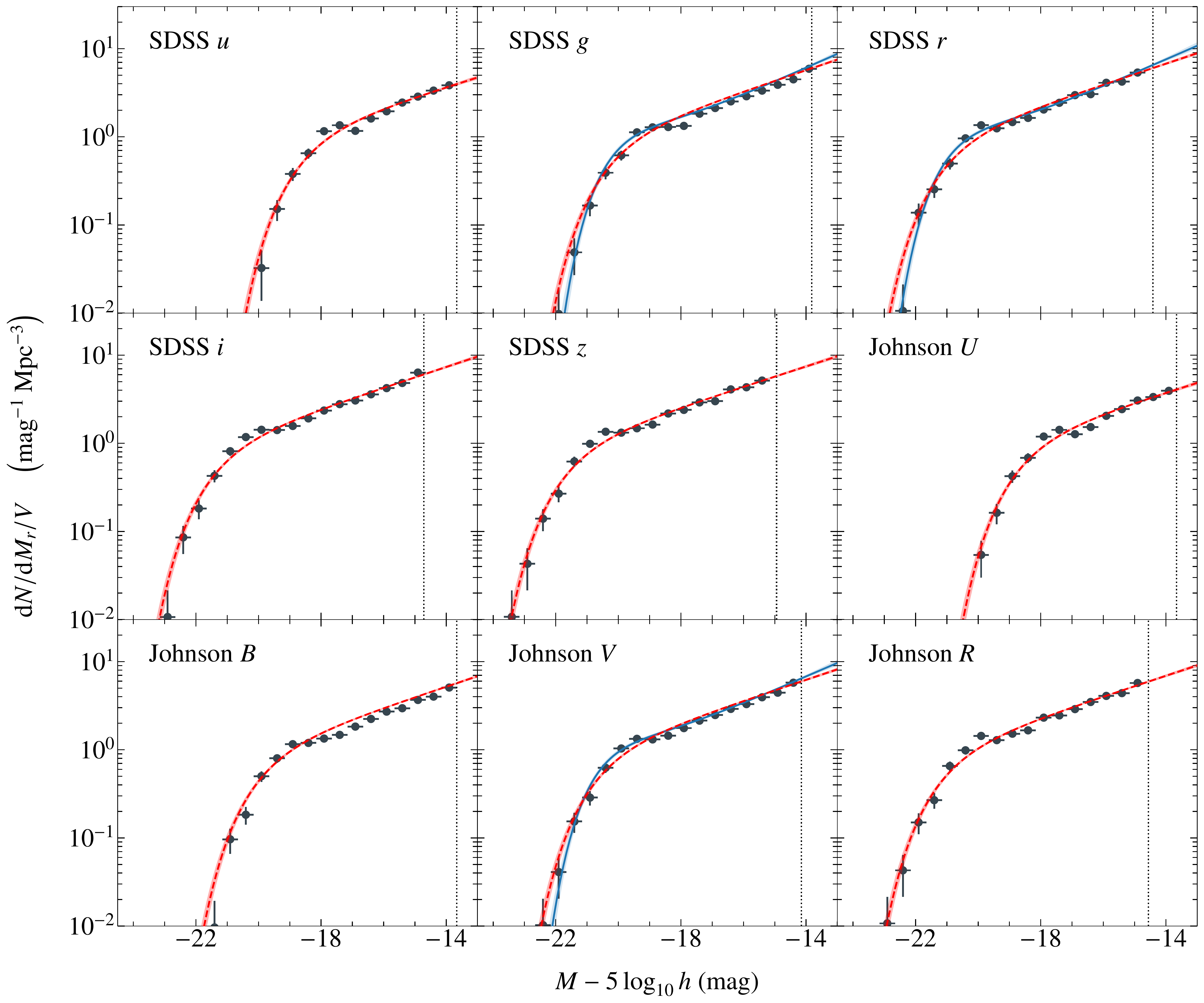}
 \caption{\CEAGLE luminosity functions for the SDSS \textit{ugriz} and Johnson \textit{UBVR} bands at $z=0$ for the intermediate mass bin (clusters having $14.46 \lesssim \log(M_{200}/\Msun) \lesssim 14.84$), with galaxies selected to have $r<r_{200}$ (the central galaxy has been removed from each cluster). Each panel shows the binned galaxy counts with Poissonian error bars as black circles, and the single Schechter fit as a red dashed line with $1\sigma$ error shaded area. For the data that shows a positive evidence for a double Schechter fit, the latter is plotted as solid blue line, while the black vertical dotted line marks the limiting magnitude in the different bands.}
 \label{fig_LF1}
\end{figure*}
\begin{figure*}
\centering
 \includegraphics[keepaspectratio, width=\textwidth]{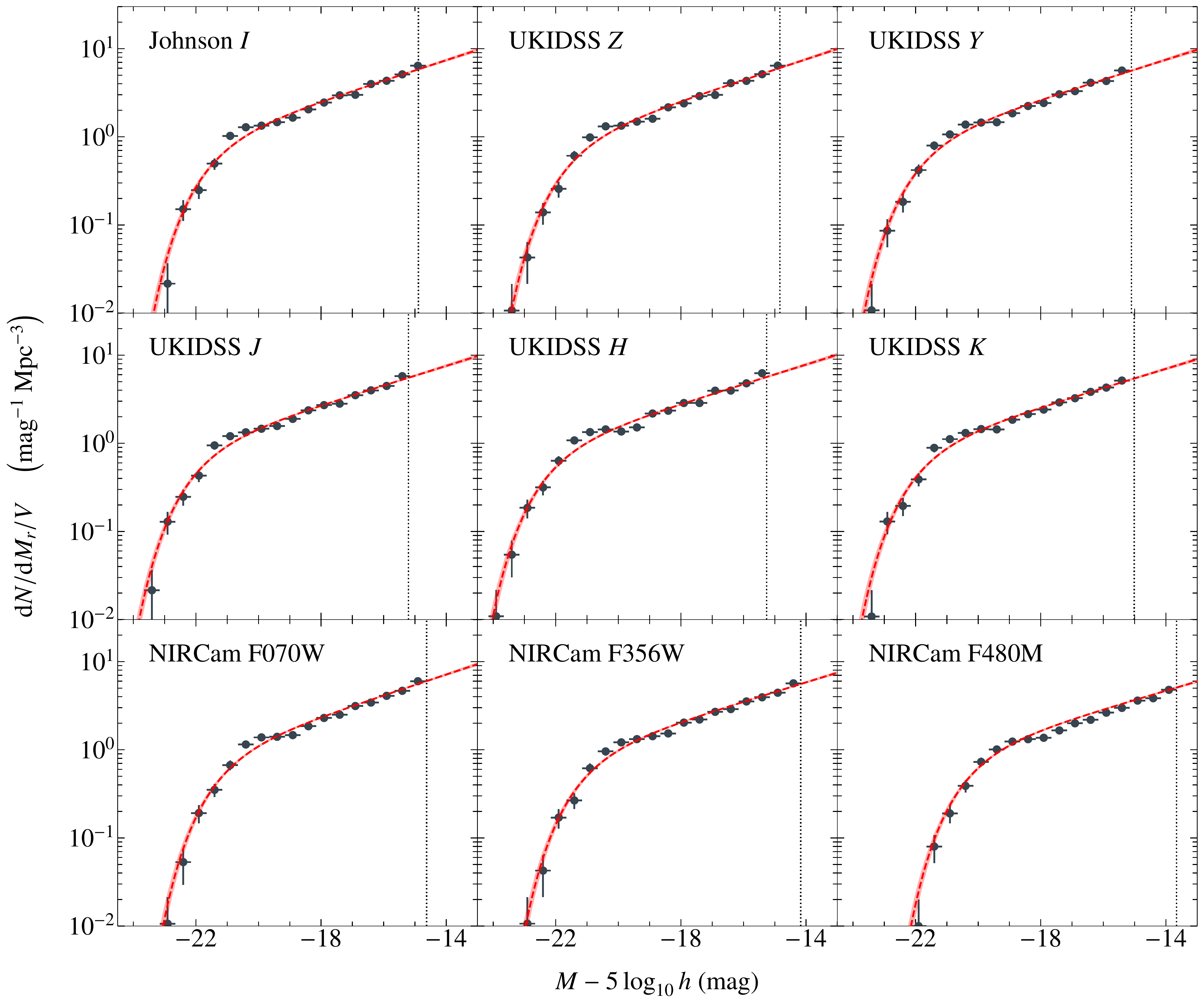}
 \caption{\CEAGLE luminosity functions for the Johnson $I$, UKIDSS \textit{ZYJHK} bands, and three of the NIRCam general purpose bands, F070W, F356W and F444W. As in figure~\ref{fig_LF1}, only the intermediate mass bin is shown.}
 \label{fig_LF2}
\end{figure*}

We present here the \CEAGLE LFs for cluster galaxies within $r_{200}$ at $z=0$ for all the photometric bands taken into consideration in Section~\ref{sec_photometry}\footnote{In the case of NIRCam we show only the F070W, F090W, F115W, F150W, F140M, F200W, F277W, F356W, F444W bands (called "general purpose" bands in the NIRCam terminology).}. In order to increase the signal-to-noise ratio, we stacked the 30 \CEAGLE clusters in three different bins of 10 clusters each, accordingly to their halo mass at $z=0$. The halo mass ranges are: $\log(M_{200}/\Msun) \leq 14.46$, $14.46 \leq \log(M_{200}/\Msun) \leq 14.84$ and $\log(M_{200}/\Msun) \geq 14.84$. We then perform Bayesian fits for both the single and double Schechter functions, with a limiting magnitude corresponding to the stellar mass cut, $M_\star = 1.8\times 10^8~\Msun$ (see section~\ref{galSelection}).

In figures~\ref{fig_LF1} and \ref{fig_LF2} we show only the LFs and their best fits for the intermediate halo mass bin, whilst the best fit parameters of the complete sample, along with the corresponding limiting magnitudes, are reported in tables~\ref{tab_results0} to \ref{tab_results2}. In the figures, we show: the binned galaxy counts (not used for the fit computation) as circles, with Poissonian error bar; the single-Schechter best fits with $\pm 1~\sigma$ shaded areas (red lines); the double-Schechter fits with $\pm 1~\sigma$ shaded areas (solid blue line), that are plotted only in the case of statistical evidence in favor of the double Schechter function ($\Delta \mathrm{BIC}>2$). The intermediate halo mass bin is the only one that shows a positive evidence against a single Schechter fit, and solely in the SDSS $g$, $r$ and Johnson $V$ bands. All the other LFs are better described by a single Schechter function.

\begin{figure*}
\centering
 \includegraphics[keepaspectratio, width=\textwidth]{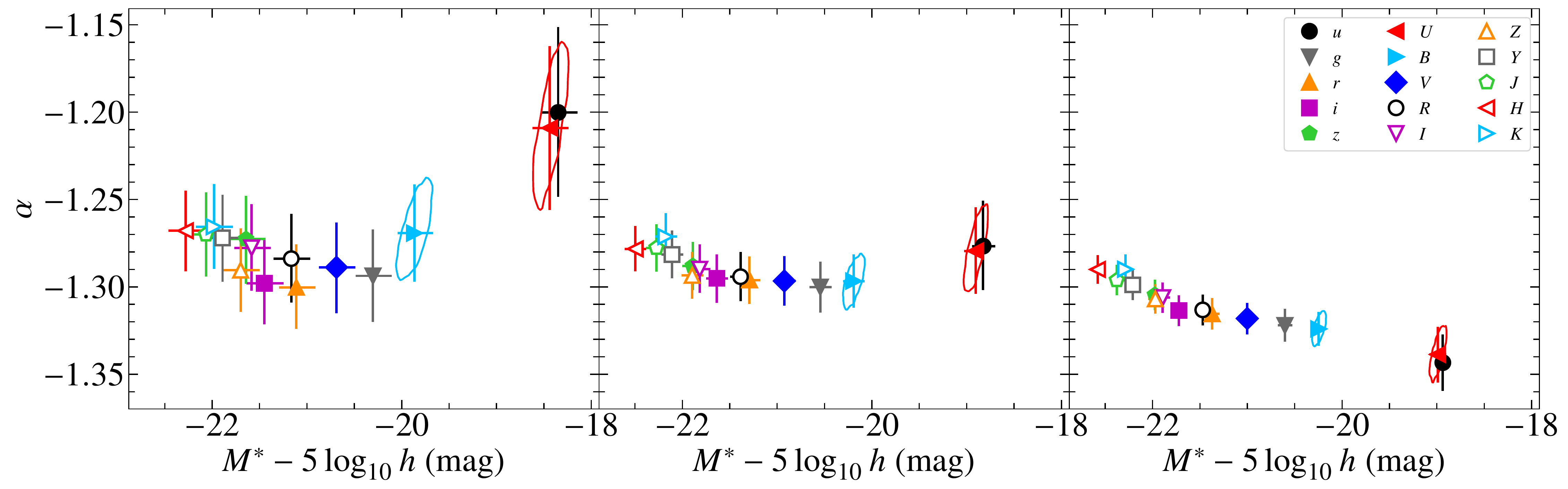}
 \includegraphics[keepaspectratio, width=\textwidth]{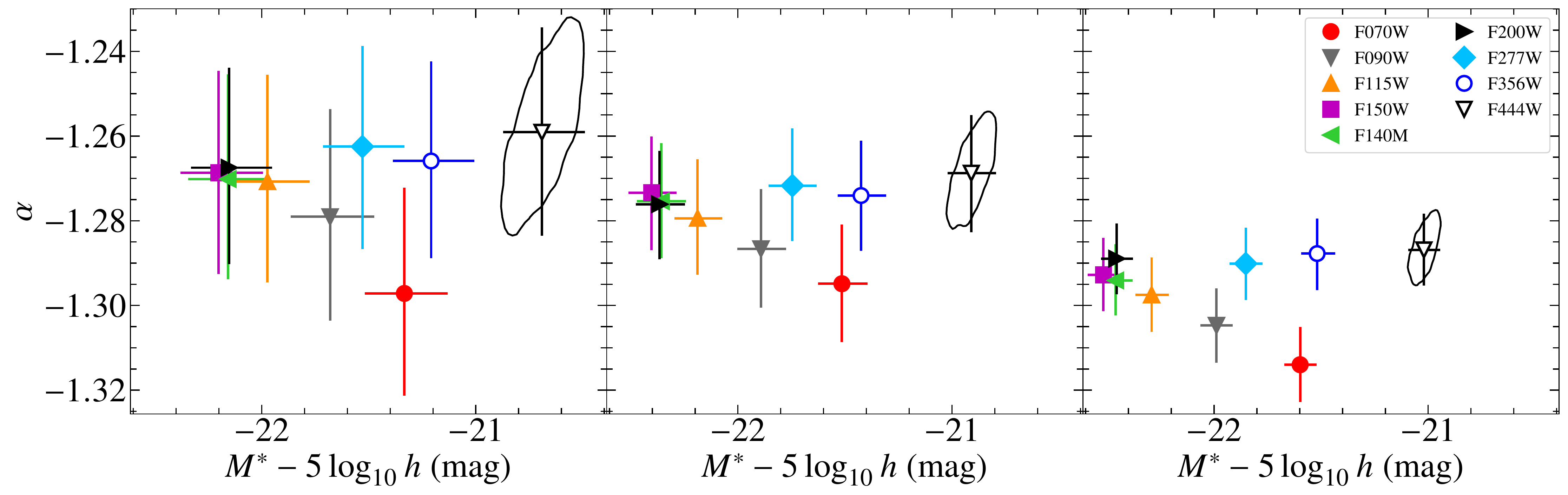}
 \caption{\CEAGLE luminosity functions' parameters $\alpha$ and $M^*$ for the low (left column), intermediate (central column) and high mass bins (right column) in the SDSS, Johnson and UKIDSS bands in the top row, while the JWST NIRCam bands are shown in the bottom row, with an narrower scale in magnitude. The error bars are the 16th and 84th percentiles of the marginalized posterior. The contours denote the $1 \sigma$ percentile of the posterior, shown only for a representative set of points.}
 \label{fig_alphaM}
\end{figure*}

We show in figure~\ref{fig_alphaM} the single-Schechter fit parameters, $\alpha$ and $M^*$, for the three halo mass bins, in different photometric bands. 2MASS filters are not included since they are very similar to the corresponding UKIDSS bands. As expected, the knee luminosity peaks around $1.5~\micron$ ($H$ and $F150W$ filters), since the virial region of the clusters is dominated by red, old galaxies (see the discussion in the previous section), whose spectra reach their maximum in the infrared; $M^*$ spans from $-19.12^{+0.21}_{-0.19}$ for the low mass bin in $u$ to $-23.43^{+0.08}_{-0.07}$ for the high mass bin in $H$. Moreover, the knee magnitude at a fixed band increases at increasing $M_{200}$, with a median difference between the low and high mass bins of $\sim 0.31$~mag in the optical and near infrared up to $0.57$~mag in the bluer bands, meaning that the more massive clusters are brighter in any band, including in the UV. This is not caused by a higher fraction of star-forming galaxies in the massive clusters, but by a larger number of luminous galaxies in the more massive clusters. Indeed, the fraction of quenched galaxies in the low, intermediate and high mass bins are $95.8_{-0.89}^{+0.44}$, $95.9_{-0.48}^{+0.24}$ and $97.6_{-0.24}^{+0.12}$, respectively\footnote{Computed with 20000 Bayesian bootstrap iterations.}, showing that the most massive bin is marginally more quenched that the lower mass ones.

The LF slope, $\alpha$, shows a modest variation, both with mass at fixed photometric band, and with photometric bands within the same mass bin. With the exception of the bluest bands for the low mass bin (upper-left panel), $\alpha$ spans the rather narrow interval $[-1.25, -1.35]$, being slightly steeper than $\alpha=-1.25\pm 0.01$ of the \EAGLE LF at $z=0$, in $r$ band (see table~\ref{tabEAGLEcont}). However, $u$ and $U$ bands do show a larger variation range, from $\alpha=1.20\pm 0.05$ to $1.34\pm 0.02$ for the low and high mass bins, respectively. This difference can be almost completely explained with the influence of the limiting magnitude on the parameters estimate. The difference between the knee and the limiting magnitudes, $M^*-M_\mathrm{f}$, quantifies how well the LF faint-end is sampled in magnitude. This quantity is a function of both photometric band and $M_{200}$; the LF is better sampled in the red bands with respect to the UV (the LF in $r$ covers 2~mags more than in $u$ for the low mass bin), and at higher masses as well, since $M_\mathrm{f}$ is constant and $M^*$ increases with mass. The former has the same effect of a more severe magnitude cut in the UV with respect to optical and IR, which produces a slightly flatter faint end (see Section~\ref{fig_EAGLEfits}). We performed a test on the $u$ band LF (black filled circles in the first row of figure~\ref{fig_alphaM}), by increasing the magnitude cut to obtain the same $M^*-M_\mathrm{f}$ independently of the mass bin, resulting in $\alpha=-1.30 \pm 0.23$ for the high mass bin, which is only marginally steeper than the low mass bin. This affects only the $u$ and $U$ band. The same tests in the other bands do not show any variation in the best fit parameters.

The vast majority of the data is well described by a single Schechter function, however the SDSS $g$, $r$ and Johnson $V$ bands of the intermediate mass bin show a barely positive ($2<\Delta \mathrm{BIC}<6$) evidence for a double Schechter function, whose best fit parameters are listed in table~\ref{tabDouble}. Thus, the \CEAGLE clusters at $z=0$ show a modest faint end upturn, with $\alpha=-1.39^{+0.04}_{-0.06}$ in all the three bands, which is steeper than all the single Schechter fits, and broadly consistent with spectroscopic observations of galaxy clusters \citep[e.g.][see the next section for a proper comparison with observations]{Agulli+2016a,Lan+2016}. This value is far from reaching the larger slopes found in photometric studies of massive clusters \citep{Popesso+2006,Drory+2009,DePropris+2018}. Finally, the normalization parameter, $\phi$, has a median of $0.995_{-0.007}^{+0.003}$ (computed with $10^5$ Bayesian bootstrap iterations), in accordance with the higher number density of galaxies in clusters with respect to the full \EAGLE simulation, which gives $\phi=8.16^{+0.30}_{-0.29} \times 10^{-3}$ (see Table~\ref{tabEAGLE}).

\subsection{Comparison with observations}\label{sec_observations}
In this section, we compare the \CEAGLE LFs with observations. We can divide the literature into two broad samples: observations of single clusters, typically at low redshift, where a photometric and spectroscopic analysis can be performed on a large number of galaxies, and stacked samples, where the clusters are typically stacked together depending on their mass or redshift. Each work has of course its own cluster selection procedures, different  completeness limits, and different apertures in which galaxies are considered part of the cluster. To ensure a consistent comparison between observations and simulations, for each observational work we re-compute the \CEAGLE luminosity functions by taking into account the observation's aperture (either fixed or as a function of $r_{200}$), its redshift, limiting magnitude, and cluster mass selection. %In addition, unless explicitly mentioned, the adopted fits are single Schechter.

\subsubsection{Comparison with individual clusters}\label{sec_individual_clusters}

\begin{figure}
\centering
 \includegraphics[keepaspectratio, width=0.4\textwidth, trim=50 0 0 0]{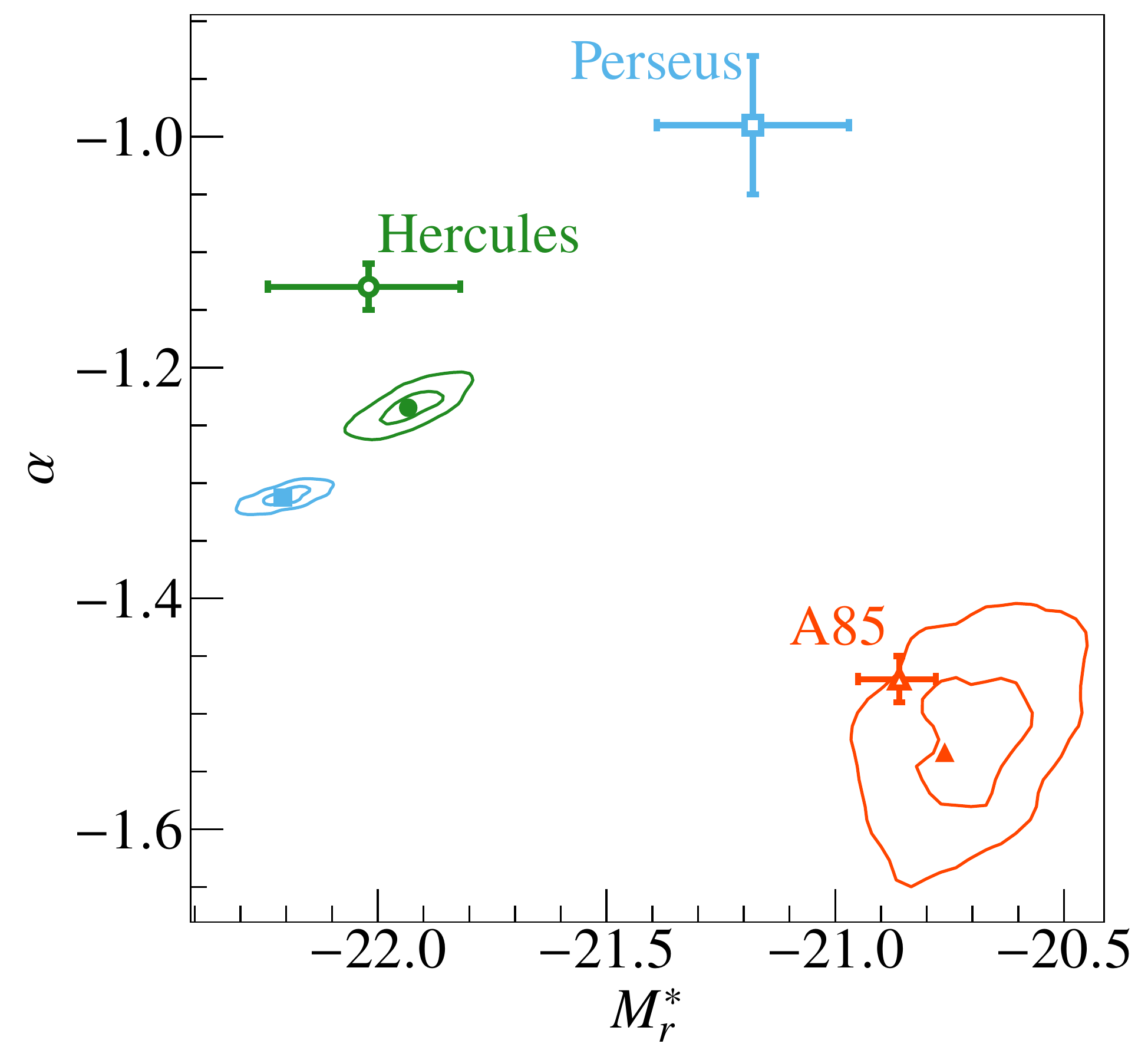}
 \caption{Comparison between luminosity functions of individual simulated clusters with three observed clusters: Abell 85 \citep{Agulli+2016a}, Hercules \citep{Agulli+2017} and Perseus \citep{Aguerri+2020}. The thick empty symbols denote the observed clusters, while the filled symbols in the same colour as the empty ones mark the corresponding simulated cluster with $M_{200}$ close to the observed one, the contours denote the $0.5$ and $1 \sigma$ percentiles of the posterior. The observations have been converted to the \CEAGLE cosmology.}
 \label{fig_obs_single}
\end{figure}

In figure~\ref{fig_obs_single}, we plot the LF parameters of three local clusters, Abell 85 \citep{Agulli+2016}, Hercules \citep{Agulli+2017} and Perseus \citep{Aguerri+2020}, which have been subject of deep spectroscopic observations. For the comparison to each of the observed LFs, we stacked the \CEAGLE clusters of mass $M_{200}$ within $\pm 0.2~\mathrm{dex}$ from the observed cluster mass, and adopting similar $M_\mathrm{f}$ and aperture of the observations. For each observation's redshift, we selected from the simulations the halo containing the main progenitor of the central galaxy at $z=0$.

Abell 85 (empty orange triangle in figure~\ref{fig_obs_single}) is a relaxed cluster at $z=0.055$, with mass $M_{200} = 2.5\times 10^{14}~\Msun$. \citet{Agulli+2016} observed Abell 85 within $r_{200}$, with a limiting magnitude of -16~mag in the adopted cosmology, finding that the LF shows an upturn at its faint end, therefore well fitted by a double Schechter function. We stacked clusters CE-6, 7, 9, 10 and 11 at $z=0.073$ (the closest redshift in the simulations' output) with $M_\mathrm{f} = -16.22$. The resulting fit shows evidence for a double Schechter function ($\Delta \mathrm{BIC}=3.6$) with $M_r^* = -20.76_{-0.27}^{+0.24}$ and $\alphaf = -1.54^{+0.11}_{-0.14}$. In figure~\ref{fig_obs_single}, we plot the double Schechter LF's parameters as a filled orange triangle. The parameters are in agreement, within the observational error, with the observed values.

The Hercules cluster, shown as a green empty circle, has $M_{200} = 4\times 10^{14}~\Msun$ at $z=0.0366$. This particular cluster is still in his collapsing phase and its LF, calculated within $1.3\times r_{200}$, has a shallower faint end then that of field galaxies \citep{Agulli+2017}. We stacked clusters CE-12 to 17 at $z=0.036$, with $M_\mathrm{f} = -16.72$, corresponding to the last bin of \citet{Agulli+2017} LF, obtaining a single-Schechter fit with $M_r^* = -21.95_{-0.14}^{+0.15}$ and $\alpha = -1.24\pm 0.03$. In figure~\ref{fig_obs_single}, we plot the corresponding fit as filled green circles. \CEAGLE clusters agree well with Hercules' knee magnitude, although showing about 0.1 larger faint-end slope.

Finally, the Perseus cluster is shown as blue empty square, having a mass of $1.2\times 10^{15}~\Msun$ at $z=0.018$, still in its forming phase, with a flat faint end \citep{Aguerri+2020}. In this case, we employed CE-22, 24, 25, 26 and 27, with an aperture of $1.4\times r_{200}$ and $M_\mathrm{f} = -16.07$ at redshift $z=0.036$. The resulting single-Schechter fit has $M_r^* = -22.23 \pm 0.11$ and $\alpha = -1.31\pm 0.015$, marked by the filled blue square. These values are in stark contrast with the observations, the simulations having one magnitude brighter knee and relatively large faint-end slope. This discrepancy is likely due to selection effects in compiling the \CEAGLE sample. We recall that the sample of simulated clusters includes only clusters that are isolated at $z=0$, without any structure having a larger mass within 20 virial radii, which is not the type of environment in which the Perseus cluster resides. The larger knee luminosity and faint-end slope of the simulated clusters possibly indicate that Perseus formed relatively recently, and that it is still far from relaxation. Observationally, the X-ray isophotes of the Perseus clusters are elongated \citep{Ettori+1998}, an X-ray substructure has been detected \citep{Mohr+1993} and the temperature map presents a substructure that have been interpret as the footprint of recent mergers with other clusters or groups \citep{Schwarz+1992,Ettori+1998,Furusho+2001}. Together with the observed strong spatial segregation of galaxy morphologies \citep{Andreon1994}, this could  indicate that the Perseus cluster is dynamically unrelaxed. This could bias the determination of its virial mass and, therefore, the comparison with simulations.

\subsubsection{Comparison with observed samples of clusters}\label{sec_stacked_clusters}
We compare here the \CEAGLE LFs with studies that push the cluster LFs investigation at high redshift by stacking clusters in bins of mass and redshift. As we did in section~\ref{sec_individual_clusters}, we stack simulated clusters with mass within the ranges of the observations. However, we do not sample the clusters within the redshift interval of observations, but we choose the output closest to the median redshift. We also verified that the simulated LFs do not change substantially over time.

\begin{figure}
\centering
 \includegraphics[keepaspectratio, width=0.4\textwidth, trim=50 0 0 0]{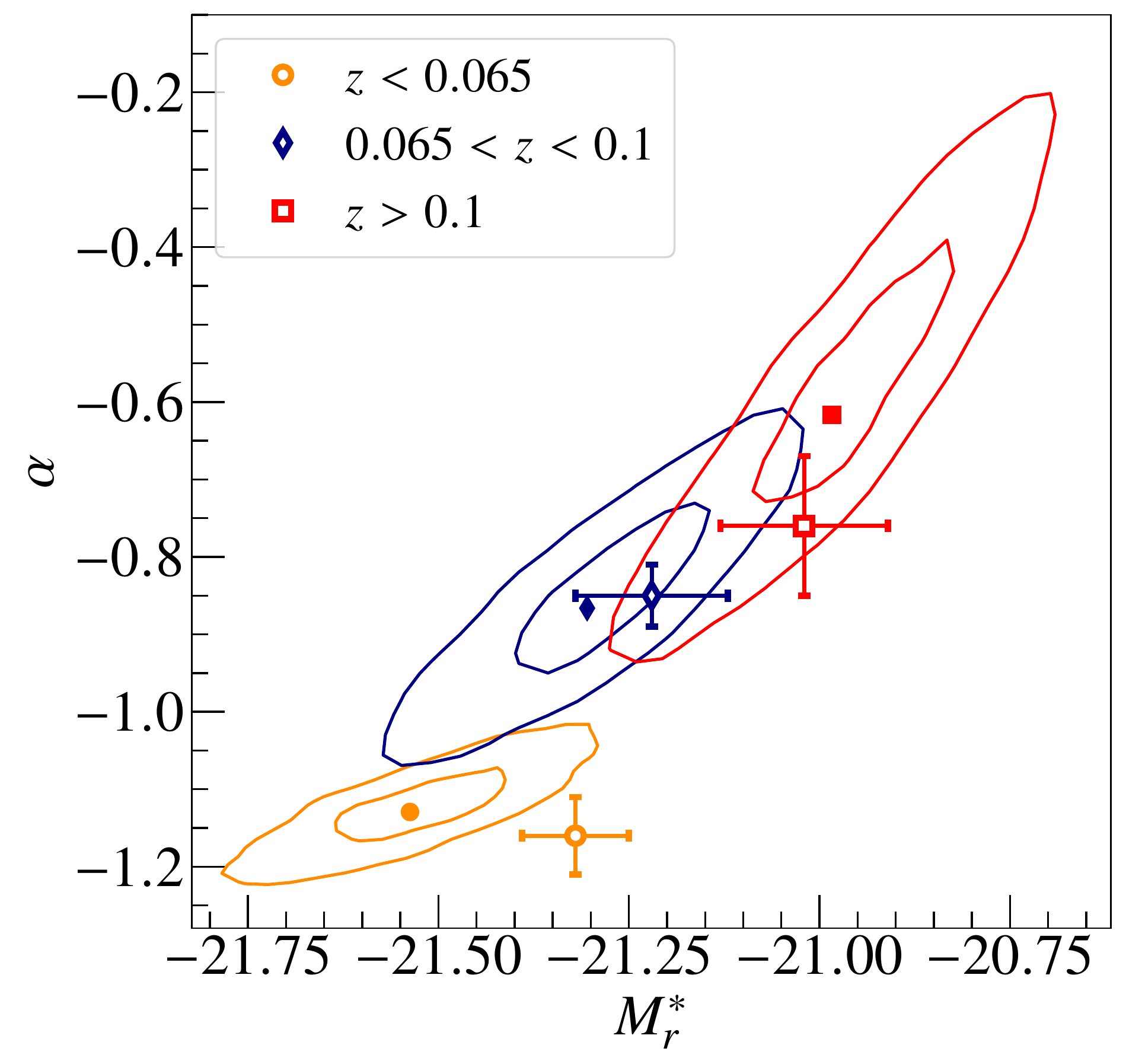}
 \caption{Stacked simulated and observed cluster fit parameters comparison. The observed cluster values are taken from \citet{OMill+2019}, where they are binned in redshift, and here represented by thick empty symbols. The observations have been converted to the \CEAGLE cosmology. Symbols and contours denotes the same quantities as in figure~\ref{fig_obs_single}.}
 \label{fig_obs_stacked}
\end{figure}

We compare our results with the low X-ray luminosity clusters of \citet{OMill+2019}. This is a sample of 45 galaxy clusters observed with ROSAT and being fainter than $7\times 10^{43}~\ergs$ in X-ray. These were later observed within SDSS, and stacked in three bins of redshift ($z<0.065$, $z>0.1$ and intermediate values), with a maximum redshift $z=0.2$. \citet{OMill+2019} adopted a distance of $1~\Mpc$ from the X-ray centroid as necessary condition for cluster membership. We plot the best fit parameters for the observed LFs in figure~\ref{fig_obs_stacked} as empty circle, square and diamond. To compute the simulated LF equivalent, we adopted the clusters of the least massive bin, at the snapshots $z=[0.036,0.073,0.155]$, with $M_\mathrm{f}=[-18.22, -19.47, -19.97]$, plotted as filled circles, squares and diamond in the figure. The \CEAGLE LFs are in agreement with the observed stacked clusters at every redshift, although we note that at high redshift the combination of the severe magnitude limit and low galaxy counts produces rather large uncertainties in $\alpha$.

\begin{figure}
\centering
 \includegraphics[keepaspectratio, width=0.4\textwidth, trim=50 0 0 0]{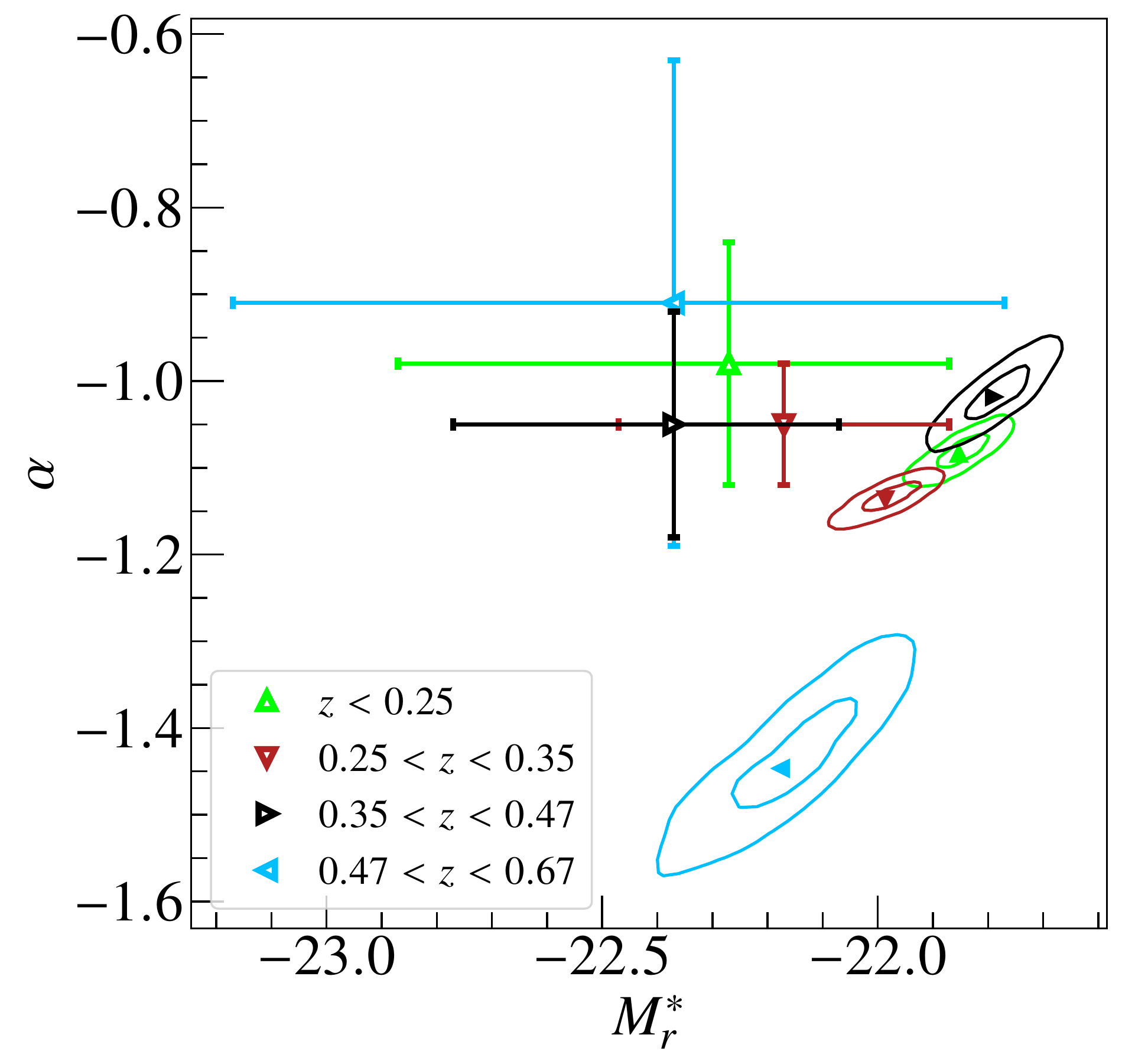}
 \caption{Same as figure~\ref{fig_obs_stacked}, where the observed clusters are taken from \citet{Ricci+2018} (empty symbols).}
 \label{fig_obs_stacked2}
\end{figure}

\citet{Ricci+2018} studied the LFs of the XXL survey clusters, a sample of 142 clusters from $z=0.03$ to 1.06 with $10^{13}<M_{500}/\Msun < 5 \times 10^{14}$, obtained by overlapping the XXL survey with the W1 field of the Canada–France–Hawaii Telescope Legacy Survey (CFHTLS). They stacked the clusters according to richness (for clusters at $z<0.67$) or redshift (for richness larger than 6), with redshift bins bounded by the values [0, 0.25, 0.35, 0.47, 0.67]. We show in figure~\ref{fig_obs_stacked2} their fit parameters in the case of $z$-stacking as empty triangles, not including the $z>0.67$ bin due to the large error in $\alpha$. We selected the \CEAGLE snapshots closest to the right bound of each redshift bin. We used their figure~8 to compute the limiting magnitude, obtaining $M_\mathrm{f} = [-18.32, -17.93, -18.96, -20.07]$ for the snapshots at $z=[0.247, 0.352, 0.474, 0.7]$, while the adopted clusters are CE-0 to 21, plus CE-23 and 25 at $z=0.247$, in addition to CE-27 at $z=0.352$, while we used all of \CEAGLE except CE-26 and 29 at $z=0.474$, and CE-22 and 26 at $z=0.7$. The observed galaxies are inside $r_{500}$, and their luminosity is in $r'$ band. We adopted the same aperture and cluster mass selection, but used $r$ band photometry data, noting that the differences with the $r'$ band are smaller than $0.06$~mags for galactic objects. This is relatively small when compared with the uncertainties that affect the observed $M_r^*$. The results are shown in figure~\ref{fig_obs_stacked2} as filled triangles. All the \CEAGLE LFs show a rather flat faint end, except at the highest $z$, without a large variation of the best-fit parameters, and they broadly agree with \citet{Ricci+2018} results, although the $0.35<z<0.47$ bin shows a fainter knee for the simulated galaxies. The severe magnitude cut at $z=0.7$ and our wide flat prior in $M^*_r$ produces a higher knee luminosity and a much steeper faint end, with larger errors. By employing a less permissive Gaussian prior\footnote{The adopted prior is a multidimensional Gaussian having $[-1, -22, 2]$ as mean for $\alpha$, $M^*_r$ and $\phi$, with standard deviation $[0.2, 0.5, 1]$.} we obtain $M^*_r=-21.85^{+0.18}_{-0.17}$ and $\alpha=-1.16 \pm 0.10$, while the other best fits parameters in figure~\ref{fig_obs_stacked2} do not show any significant variation when the Gaussian prior is employed. We finally note that the stacking in mass allow for smaller uncertainties in both $\alpha$ and $M^*_r$ with respect to the observation, contrary to the comparison with \citet{OMill+2019}.

\section{Summary and conclusions}\label{sec_conclusions}
In this work, we developed a method for calculating the rest-frame luminosities and spectra of simulated galaxies of the \CEAGLE simulation project, a suite of zoom-in simulations of 30 galaxy clusters in the mass range of $10^{14} < M_{200}/\Msun < 10^{15.4}$ at $z=0$ \citep{Bahe+2017,Barnes+2017}. The simulations make use of the \EAGLE AGNdT9 model \citep{Schaye+2015,Crain+2015}, and their high-resolution volumes extend up to $10\times r_{200}$ from the cluster centre for 24 clusters \citep[forming the \textit{Hydrangea} sample,][]{Bahe+2017}, whilst all 30 clusters have been simulated in high-resolution volumes of radius $\geq 5\times r_{200}$ \citep{Barnes+2017}. The clusters are selected to be at least at a minimum distance of either $30~\Mpc$ or $20 \times r_{200}$ from other clusters of higher mass, at $z=0$.

We computed the AB magnitudes of the simulated galaxies in the SDSS \textit{ugriz}, Johnson-Cousins \textit{UBVRI}, UKIDSS \textit{JHKYZ}, 2MASS \textit{JHK$_s$} and NIRCam photometric systems, adopting the SSP model based on the E-MILES spectral libraries \citep{Vazdekis+2016}. We resampled recent star formation to alleviate large flux fluctuations (specially in the UV continuum) due to the stochastic conversion of gas particles into stellar particles in the numerical model. We have taken into account dust obscuration by employing a modified version of the GD model used in \EAGLE \citep{Trayford+2015}, where, in our case, the dust attenuation linearly depends on the star-forming gas metallicity and projected gas surface density instead of total gas mass, yielding good results also for massive central galaxies. Galaxy magnitudes are computed in a series of concentric spherical apertures up to $100~\kpc$.

We first compared our results with data from the Ref \EAGLE simulation. Doing this, we selected $30~\kpc$ as fiducial aperture throughout the all analysis. This choice yields stellar masses comparable to those recovered within a projected circular aperture of the Petrosian radius \citep{Schaye+2015}.
The $g-r$ colour-stellar mass diagram at $z=0.1$ (figure~\ref{color_mass}) shows a well populated blue cloud and defined red sequence when all galaxies in the simulated high-resolution region are considered. The blue cloud is absent within $r_{200}$, in agreement with the high fractions of quenched galaxies found in \citet{Bahe+2017}, although few star forming galaxies having $\Mstar > 10^{10.5}~\Msun$ are still present. As in the Ref \EAGLE simulation, we note that at $\Mstar <10^9~\Msun$, the \CEAGLE clusters inside $10\times r_{200}$ contain an unrealistically high number of red dwarfs, due mostly to numerical resolution \citep{Trayford+2015}. We fit the red sequence of both \CEAGLE and \EAGLE with a broken power-law, to take into account its flattening at high stellar mass.
At $\Mstar < 10^{10.4}~\Msun$ both red sequences show that the $g-r$ colour increases with increasing mass, and have similar dispersion around the mean. A higher masses, the \EAGLE red sequence flattens, while the \CEAGLE one still show an increase of $g-r$ with mass, becoming $0.05$~dex redder than \EAGLE at $10^{11.5}~\Msun$. As expected, the massive end of the red sequence is populated by the 30 central galaxies at $z=0.1$, with $g-r \gtrsim 0.8$, due to their evolution in an high density environment. 

We adopted Bayesian inference to compute the LFs best fits with both a single and double Schechter function. The double Schechter function was employed to account for the faint-end steepening of clusters' LFs observed in photometric and spectroscopic surveys. We selected the best model describing the data with the aid of the Bayesian Information Criterion (section~\ref{sec_lumFunc}). The main advantages of this method are a consistent error estimation and not having to bin the data in magnitude. We tested our galaxy selection and LF computation against the \EAGLE data at $z=0.1$ (section~\ref{sec_code_validation}). The comparison show a mild dependence of the LF best-fit parameters on the adopted magnitude cut (figure~\ref{fig_EAGLEfits}), but the result is broadly consistent with \citet{Trayford+2015}, and the agreement is within their error estimate. We note, however, that the method we used yields lower errors on the best-fit parameters.

We computed the luminosity functions of the \CEAGLE clusters' virial region ($r<r_{200}$) at $z=0$ (figures~\ref{fig_LF1} and \ref{fig_LF2}), by stacking the 30 clusters in 3 mass bins, having $\log (M_{200}/\Msun)\lesssim 14.46$, $14.46 \lesssim \log (M_{200}/\Msun) \lesssim 14.84$, and $\log (M_{200}/\Msun) \gtrsim 14.84$. The multi-colour LFs show that the knee luminosity increases from UV to IR, reaching its maximum in the $H$ and F150W bands, in addition to increasing with halo mass (figure~\ref{fig_alphaM}), as expected for a sample of old, quenched galaxies at the center of massive clusters. The faint end slope $\alpha$ of the single Schechter fit is fairly independent from mass and photometric band, and constrained in the $[-1.25, -1.35]$ interval, steeper that the field LF (for \EAGLE $\alpha = -1.25$). While the vast majority of the LFs are well described by a single Schechter function, the double Schechter fit statistically more relevant for the intermediate mass bin in the SDSS $g$, $r$ and Johnson $V$ bands. In this case, the LF present a modest faint end upturn, $\alpha=-1.39^{+0.04}_{-0.07}$, broadly consistent with the spectroscopic observations of galaxy clusters \citep{Agulli+2016,Lan+2016}.

We compared the \CEAGLE $r$-band luminosity functions with low and high redshift observations. For each set of observations, we matched, as much as possible, the projected size of the observed region of the sky, the magnitude limit of the observations, the mass(es) of the observed cluster(s) and the (average) redshift of the observations.
The simulations are able to reproduce the LF of individual, nearby clusters, such as Hercules and Abell 85, the latter characterised by the upturn at the faint-end (figure~\ref{fig_obs_single}). Different is the case of Perseus cluster, with the best-fit parameters of the stacked LF of similar mass \CEAGLE clusters falling way off ($> 4\sigma$) the observed values. We argued that the discrepancy is due to selection effects in the \CEAGLE sample, where fairly isolated cluster haloes have been resimulated, in contrast with the rapidly evolving environment and the non relaxed state of Perseus cluster.

The \CEAGLE simulations reproduce the LFs of \citet{OMill+2019} up to $z=0.1$. However, we note that the small number of matching simulated clusters, combined with the strong magnitude limits of the observations, produces rather large error bars in our analysis (figure~\ref{fig_obs_stacked}). The simulations produce LFs at higher redshift that agree, within the errors, with the data from the XXL survey \citep{Ricci+2018} in three of the four redshift bins observed.

In a series of companion articles, we will study the evolution of the LF in the clusters' virial regions and outskirts.

\section*{Data Availability}
The data presented in the figures are available in the article and in its online supplementary material. The raw simulation data can be obtain through the Hydrangea python library\footnote{\url{https://hydrangea.readthedocs.io}.}. The galaxy magnitudes will be shared on reasonable request to the corresponding author. The E-MILES SSP models are publicly available at the MILES website (http://miles.iac.es).

\section*{Acknowledgements}
This python packages Numpy\footnote{https://numpy.org} \citep{Harris+2020}, Astropy\footnote{http://www.astropy.org} \citep{AstropyCollaboration+2013, AstropyCollaboration+2018}, \textit{corner.py} \citep{Foreman-Mackey2016}, \textsc{emcee} \citep{Foreman-Mackey+2013}, matplotlib \citep{Hunter2007}, CuPy \citep{Okuta+2017} have been used in this research. AN thanks Lisa Nortmann, Hannu Parviainen and Stefano Andreon for their valuable advice on MCMC; Ignacio Ferreras and James Angthopo for useful discussion on stellar spectra. CDV is grateful to Alexandre Vazdekis and Jes\'us Falc\'on Barroso for providing the E-MILES stellar spectral library before being made public, and for their help during the writing of the first version of the analysis code used in this work. CDV is also thankful to Michael Beasley for thorough discussions that helped the design of the code. AN and CDV are supported by the Spanish Ministry of Science and Innovation (MICIU/FEDER) through research grant PGC2018-094975-C22. CDV acknowledges support from MICIU through grant RYC-2015-18078. YMB acknowledges funding from the EU Horizon 2020 research and innovation programme under Marie Sk\l{}odowska-Curie grant agreement 747645 (ClusterGal) and the Netherlands Organisation for Scientific Research (NWO) through VENI grant number 639.041.751. This research made use of computing time available on the high-performance computing systems, Deimos/Diva, of the Instituto de Astrofisica de Canarias. The authors thankfully acknowledges the technical expertise and assistance provided by the Spanish Supercomputing Network (RES). The Hydrangea simulations were in part performed on the German federal maximum performance computer “HazelHen” at the maximum performance computing centre Stuttgart (HLRS), under project GCS-HYDA / ID 44067 financed through the large-scale project “Hydrangea” of the Gauss Center for Supercomputing. Further simulations were performed at the Max Planck Computing and Data Facility in Garching, Germany. This work used the DiRAC@Durham facility managed by the Institute for Computational Cosmology on behalf of the STFC DiRAC HPC Facility (www.dirac.ac.uk). The equipment was funded by BEIS capital funding via STFC capital grants ST/K00042X/1, ST/P002293/1, ST/R002371/1 and ST/S002502/1, Durham University and STFC operations grant ST/R000832/1. DiRAC is part of the National e-Infrastructure.

\bibliographystyle{mnras}
\bibliography{aaa_new}

\appendix
\section{Best fit parameters}
In this appendix we present the best fit parameters tables for all the photometric bands and mass bins studied in this paper.
\include{tab_fits1Sc}

\include{tab_fits2Sc}

\bsp
\label{lastpage}
\end{document}

%% file: tab_fits1sc.tex
\begin{table}
    \caption{Single Schechter best fit parameters for the low mass bin stacked samples' LFs within $r_{200}$ at $z=0$. Values and errors are the median and the $1\sigma$ percentile of the posterior parameter distribution. The columns refer to: (1) filter name, (2) number of galaxies selected, (3) knee magnitude (in mags), (4) faint-end slope, (5) normalization (in $h^3 \cMpc^{-3} \mathrm{mag}^{-1}$), (6) limiting magnitude (in mags). A machine readable version is present in the online material, along with a plot of the posterior for each fit.}\label{tab_results0}
\begin{tabular}{cccccccc}
\toprule
Filter &        $N$ &      $M^*$   &           $\alpha$    &                 $\phi$     &      $M_\mathrm{f}$ \\
       %&            &     (mag)    &                       &$(h^3 \cMpc^{-3} \mathrm{mag}^{-1})$ & (mag) \\
    (1) & (2) & (3) & (4) & (5) & (6) \\
\midrule
F070W            & 1102& $-22.18^{+ 0.19}_{- 0.20}$ & $ -1.30^{+ 0.02}_{- 0.02}$ &$  1.01^{+ 0.17}_{- 0.15}$ &          -15.47 \\[1.3ex]
F090W            & 1058& $-22.53^{+ 0.18}_{- 0.21}$ & $ -1.28^{+ 0.03}_{- 0.02}$ &$  1.04^{+ 0.17}_{- 0.15}$ &          -15.82 \\[1.3ex]
F115W            & 1054& $-22.82^{+ 0.18}_{- 0.20}$ & $ -1.27^{+ 0.03}_{- 0.02}$ &$  1.03^{+ 0.17}_{- 0.14}$ &           -16.0 \\[1.3ex]
F140M            & 1056& $-23.00^{+ 0.19}_{- 0.20}$ & $ -1.27^{+ 0.02}_{- 0.02}$ &$  1.01^{+ 0.16}_{- 0.15}$ &          -16.11 \\[1.3ex]
F150W            & 1056& $-23.05^{+ 0.18}_{- 0.21}$ & $ -1.27^{+ 0.02}_{- 0.02}$ &$  1.02^{+ 0.16}_{- 0.15}$ &          -16.14 \\[1.3ex]
F200W            & 1080& $-23.00^{+ 0.18}_{- 0.20}$ & $ -1.27^{+ 0.02}_{- 0.02}$ &$  1.00^{+ 0.16}_{- 0.14}$ &          -15.96 \\[1.3ex]
F277W            & 1051& $-22.38^{+ 0.18}_{- 0.20}$ & $ -1.26^{+ 0.02}_{- 0.02}$ &$  1.03^{+ 0.16}_{- 0.15}$ &          -15.44 \\[1.3ex]
F444W            & 1023& $-21.54^{+ 0.18}_{- 0.20}$ & $ -1.26^{+ 0.02}_{- 0.02}$ &$  1.04^{+ 0.16}_{- 0.15}$ &          -14.68 \\[1.3ex]
$U$              &  647& $-19.28^{+ 0.18}_{- 0.20}$ & $ -1.21^{+ 0.05}_{- 0.05}$ &$  1.55^{+ 0.30}_{- 0.27}$ &          -14.51 \\[1.3ex]
$B$              &  995& $-20.71^{+ 0.18}_{- 0.20}$ & $ -1.27^{+ 0.03}_{- 0.03}$ &$  1.20^{+ 0.19}_{- 0.18}$ &          -14.51 \\[1.3ex]
$V$              & 1071& $-21.54^{+ 0.18}_{- 0.20}$ & $ -1.29^{+ 0.03}_{- 0.03}$ &$  1.07^{+ 0.18}_{- 0.16}$ &           -15.0 \\[1.3ex]
$R$              & 1071& $-22.01^{+ 0.19}_{- 0.20}$ & $ -1.28^{+ 0.03}_{- 0.03}$ &$  1.07^{+ 0.18}_{- 0.15}$ &           -15.4 \\[1.3ex]
$I$              & 1069& $-22.43^{+ 0.18}_{- 0.20}$ & $ -1.28^{+ 0.02}_{- 0.02}$ &$  1.07^{+ 0.17}_{- 0.16}$ &          -15.74 \\[1.3ex]
2MASS   $H$      & 1041& $-23.10^{+ 0.18}_{- 0.20}$ & $ -1.26^{+ 0.02}_{- 0.02}$ &$  1.05^{+ 0.17}_{- 0.15}$ &         -16.21 \\[1.3ex]
2MASS   $J$      & 1057& $-22.89^{+ 0.18}_{- 0.19}$ & $ -1.27^{+ 0.02}_{- 0.02}$ &$  1.03^{+ 0.16}_{- 0.15}$ &         -16.04 \\[1.3ex]
2MASS \textit{Ks}& 1031& $-22.85^{+ 0.18}_{- 0.20}$ & $ -1.26^{+ 0.02}_{- 0.02}$ &$  1.04^{+ 0.16}_{- 0.15}$ &         -15.96 \\[1.3ex]
UKIDSS  $H$      & 1082& $-23.13^{+ 0.18}_{- 0.20}$ & $ -1.27^{+ 0.02}_{- 0.02}$ &$  1.01^{+ 0.15}_{- 0.14}$ &         -16.11 \\[1.3ex]
UKIDSS  $J$      & 1055& $-22.91^{+ 0.18}_{- 0.20}$ & $ -1.27^{+ 0.02}_{- 0.02}$ &$  1.03^{+ 0.16}_{- 0.15}$ &         -16.05 \\[1.3ex]
UKIDSS  $K$      & 1055& $-22.83^{+ 0.19}_{- 0.20}$ & $ -1.27^{+ 0.02}_{- 0.02}$ &$  1.01^{+ 0.16}_{- 0.15}$ &         -15.86 \\[1.3ex]
UKIDSS  $Y$      & 1057& $-22.74^{+ 0.18}_{- 0.20}$ & $ -1.27^{+ 0.02}_{- 0.02}$ &$  1.04^{+ 0.17}_{- 0.15}$ &         -15.95 \\[1.3ex]
UKIDSS  $Z$      & 1117& $-22.55^{+ 0.19}_{- 0.20}$ & $ -1.29^{+ 0.02}_{- 0.02}$ &$  1.00^{+ 0.16}_{- 0.15}$ &        -15.68 \\[1.3ex]
  $u$            &  628& $-19.20^{+ 0.18}_{- 0.21}$ & $ -1.20^{+ 0.05}_{- 0.05}$ &$  1.59^{+ 0.31}_{- 0.28}$ &          -14.51 \\[1.3ex]
  $g$            & 1067& $-21.15^{+ 0.18}_{- 0.20}$ & $ -1.29^{+ 0.03}_{- 0.03}$ &$  1.07^{+ 0.18}_{- 0.16}$ &          -14.67 \\[1.3ex]
  $r$            & 1102& $-21.96^{+ 0.18}_{- 0.20}$ & $ -1.30^{+ 0.02}_{- 0.02}$ &$  1.00^{+ 0.16}_{- 0.15}$ &         -15.26 \\[1.3ex]
  $i$            & 1107& $-22.30^{+ 0.19}_{- 0.21}$ & $ -1.30^{+ 0.03}_{- 0.02}$ &$  1.00^{+ 0.17}_{- 0.15}$ &         -15.56 \\[1.3ex]
  $z$            & 1064& $-22.49^{+ 0.18}_{- 0.20}$ & $ -1.27^{+ 0.02}_{- 0.03}$ &$  1.08^{+ 0.17}_{- 0.16}$ &          -15.79 \\[1.3ex]
\bottomrule
\end{tabular}
\end{table}

\begin{table}
    \caption{Same fields as Table~\ref{tab_results0}, for the intermediate mass bin.}\label{tab_results1}
\begin{tabular}{cccccccc}
\toprule
Filter &        N &      $M^*$   &           $\alpha$    &                 $\phi$     &      $M_\mathrm{f}$ \\
    (1) & (2) & (3) & (4) & (5) & (6) \\
\midrule
F070W            & 3289& $-22.36^{+ 0.11}_{- 0.12}$ &  $ -1.29 ^{+ 0.01 }_{- 0.01}$   &$  1.01^{+ 0.10}_{- 0.09}$ &          -15.47 \\[1.3ex]
F090W            & 3190& $-22.74^{+ 0.11}_{- 0.12}$ &  $ -1.29 ^{+ 0.01 }_{- 0.01}$   &$  1.00^{+ 0.09}_{- 0.09}$ &          -15.82 \\[1.3ex]
F115W            & 3184& $-23.04^{+ 0.11}_{- 0.11}$ &  $ -1.28 ^{+ 0.01 }_{- 0.01}$   &$  0.99^{+ 0.09}_{- 0.08}$ &           -16.0 \\[1.3ex]
F140M            & 3182& $-23.20^{+ 0.11}_{- 0.12}$ &  $ -1.28 ^{+ 0.01 }_{- 0.01}$   &$  0.99^{+ 0.09}_{- 0.08}$ &          -16.11 \\[1.3ex]
F150W            & 3180& $-23.25^{+ 0.11}_{- 0.12}$ &  $ -1.27 ^{+ 0.01 }_{- 0.01}$   &$  0.99^{+ 0.09}_{- 0.08}$ &          -16.14 \\[1.3ex]
F200W            & 3265& $-23.21^{+ 0.11}_{- 0.12}$ &  $ -1.28 ^{+ 0.01 }_{- 0.01}$   &$  0.96^{+ 0.08}_{- 0.08}$ &          -15.96 \\[1.3ex]
F277W            & 3177& $-22.59^{+ 0.11}_{- 0.11}$ &  $ -1.27 ^{+ 0.01 }_{- 0.01}$   &$  0.99^{+ 0.09}_{- 0.08}$ &          -15.44 \\[1.3ex]
F444W            & 3093& $-21.76^{+ 0.11}_{- 0.11}$ &  $ -1.27 ^{+ 0.01 }_{- 0.01}$   &$  1.00^{+ 0.09}_{- 0.09}$ &          -14.68 \\[1.3ex]
$U$              & 1984& $-19.75^{+ 0.12}_{- 0.13}$ &  $ -1.28 ^{+ 0.02 }_{- 0.02}$   &$  1.15^{+ 0.14}_{- 0.13}$ &          -14.51 \\[1.3ex]
$B$              & 3028& $-21.04^{+ 0.11}_{- 0.12}$ &  $ -1.30 ^{+ 0.02 }_{- 0.02}$   &$  1.04^{+ 0.10}_{- 0.09}$ &          -14.51 \\[1.3ex]
$V$              & 3220& $-21.77^{+ 0.11}_{- 0.12}$ &  $ -1.30 ^{+ 0.01 }_{- 0.01}$   &$  1.02^{+ 0.10}_{- 0.09}$ &           -15.0 \\[1.3ex]
$R$              & 3233& $-22.23^{+ 0.11}_{- 0.12}$ &  $ -1.29 ^{+ 0.01 }_{- 0.01}$   &$  1.01^{+ 0.09}_{- 0.09}$ &           -15.4 \\[1.3ex]
$I$              & 3231& $-22.66^{+ 0.11}_{- 0.11}$ &  $ -1.29 ^{+ 0.01 }_{- 0.01}$   &$  1.00^{+ 0.09}_{- 0.09}$ &          -15.74 \\[1.3ex]
2MASS   $H$      & 3157& $-23.34^{+ 0.11}_{- 0.12}$ &  $ -1.27 ^{+ 0.01 }_{- 0.01}$   &$  0.98^{+ 0.09}_{- 0.08}$ &         -16.21 \\[1.3ex]
2MASS   $J$      & 3192& $-23.11^{+ 0.11}_{- 0.12}$ &  $ -1.28 ^{+ 0.01 }_{- 0.01}$   &$  0.99^{+ 0.09}_{- 0.09}$ &         -16.04 \\[1.3ex]
2MASS \textit{Ks}& 3121& $-23.07^{+ 0.11}_{- 0.12}$ &  $ -1.27 ^{+ 0.01 }_{- 0.01}$   &$  0.99^{+ 0.09}_{- 0.08}$ &         -15.96 \\[1.3ex]
UKIDSS  $H$      & 3274& $-23.34^{+ 0.11}_{- 0.12}$ &  $ -1.28 ^{+ 0.01 }_{- 0.01}$   &$  0.96^{+ 0.08}_{- 0.08}$ &         -16.11 \\[1.3ex]
UKIDSS  $J$      & 3182& $-23.12^{+ 0.11}_{- 0.11}$ &  $ -1.28 ^{+ 0.01 }_{- 0.01}$   &$  0.99^{+ 0.09}_{- 0.08}$ &         -16.05 \\[1.3ex]
UKIDSS  $K$      & 3179& $-23.02^{+ 0.11}_{- 0.12}$ &  $ -1.27 ^{+ 0.01 }_{- 0.01}$   &$  0.99^{+ 0.09}_{- 0.08}$ &         -15.86 \\[1.3ex]
UKIDSS  $Y$      & 3192& $-22.96^{+ 0.11}_{- 0.12}$ &  $ -1.28 ^{+ 0.01 }_{- 0.01}$   &$  0.99^{+ 0.09}_{- 0.09}$ &         -15.95 \\[1.3ex]
UKIDSS  $Z$      & 3350& $-22.74^{+ 0.11}_{- 0.12}$ &  $ -1.29 ^{+ 0.01 }_{- 0.01}$   &$  0.97^{+ 0.09}_{- 0.08}$ &        -15.68 \\[1.3ex]
  $u$            & 1928& $-19.67^{+ 0.12}_{- 0.13}$ &  $ -1.28 ^{+ 0.03 }_{- 0.03}$   &$  1.16^{+ 0.14}_{- 0.13}$ &          -14.51 \\[1.3ex]
  $g$            & 3205& $-21.39^{+ 0.11}_{- 0.12}$ &  $ -1.30 ^{+ 0.01 }_{- 0.01}$   &$  1.02^{+ 0.10}_{- 0.09}$ &          -14.67 \\[1.3ex]
  $r$            & 3288& $-22.14^{+ 0.11}_{- 0.11}$ &  $ -1.30 ^{+ 0.01 }_{- 0.01}$   &$  1.00^{+ 0.09}_{- 0.09}$ &         -15.26 \\[1.3ex]
  $i$            & 3306& $-22.48^{+ 0.11}_{- 0.12}$ &  $ -1.30 ^{+ 0.01 }_{- 0.01}$   &$  1.00^{+ 0.09}_{- 0.09}$ &         -15.56 \\[1.3ex]
  $z$            & 3226& $-22.74^{+ 0.11}_{- 0.12}$ &  $ -1.29 ^{+ 0.01 }_{- 0.01}$   &$  1.00^{+ 0.09}_{- 0.09}$ &          -15.79 \\[1.3ex]
\bottomrule
\end{tabular}
\end{table}

\begin{table}
    \caption{Same fields as Table~\ref{tab_results0}, for the high mass bin.}\label{tab_results2}
\begin{tabular}{cccccccc}
\toprule
Filter &        N &      $M^*$   &           $\alpha$    &                 $\phi$     &      $M_\mathrm{f}$ \\
    (1) & (2) & (3) & (4) & (5) & (6) \\
\midrule
F070W            & 8092& $-22.45^{+ 0.07}_{- 0.08}$ & $ -1.31 ^{+ 0.01}_{- 0.01}$          &$  0.80^{+ 0.05}_{- 0.05}$ &          -15.47 \\[1.3ex]
F090W            & 7836& $-22.84^{+ 0.08}_{- 0.08}$ & $ -1.30 ^{+ 0.01}_{- 0.01}$          &$  0.80^{+ 0.05}_{- 0.05}$ &          -15.82 \\[1.3ex]
F115W            & 7818& $-23.14^{+ 0.08}_{- 0.08}$ & $ -1.30 ^{+ 0.01}_{- 0.01}$          &$  0.79^{+ 0.05}_{- 0.05}$ &           -16.0 \\[1.3ex]
F140M            & 7824& $-23.31^{+ 0.07}_{- 0.08}$ & $ -1.29 ^{+ 0.01}_{- 0.01}$          &$  0.78^{+ 0.05}_{- 0.04}$ &          -16.11 \\[1.3ex]
F150W            & 7821& $-23.36^{+ 0.07}_{- 0.08}$ & $ -1.29 ^{+ 0.01}_{- 0.01}$          &$  0.78^{+ 0.05}_{- 0.05}$ &          -16.14 \\[1.3ex]
F200W            & 7989& $-23.30^{+ 0.07}_{- 0.08}$ & $ -1.29 ^{+ 0.01}_{- 0.01}$          &$  0.78^{+ 0.05}_{- 0.04}$ &          -15.96 \\[1.3ex]
F277W            & 7812& $-22.70^{+ 0.08}_{- 0.08}$ & $ -1.29 ^{+ 0.01}_{- 0.01}$          &$  0.78^{+ 0.05}_{- 0.04}$ &          -15.44 \\[1.3ex]
F444W            & 7597& $-21.87^{+ 0.07}_{- 0.08}$ & $ -1.29 ^{+ 0.01}_{- 0.01}$          &$  0.79^{+ 0.05}_{- 0.04}$ &          -14.68 \\[1.3ex]
$U$              & 4841& $-19.84^{+ 0.09}_{- 0.09}$ & $ -1.34 ^{+ 0.02}_{- 0.02}$          &$  0.82^{+ 0.07}_{- 0.07}$ &          -14.51 \\[1.3ex]
$B$              & 7458& $-21.09^{+ 0.07}_{- 0.08}$ & $ -1.32 ^{+ 0.01}_{- 0.01}$          &$  0.81^{+ 0.05}_{- 0.05}$ &          -14.51 \\[1.3ex]
$V$              & 7921& $-21.85^{+ 0.07}_{- 0.08}$ & $ -1.32 ^{+ 0.01}_{- 0.01}$          &$  0.81^{+ 0.05}_{- 0.05}$ &           -15.0 \\[1.3ex]
$R$              & 7942& $-22.32^{+ 0.07}_{- 0.08}$ & $ -1.31 ^{+ 0.01}_{- 0.01}$          &$  0.81^{+ 0.05}_{- 0.05}$ &           -15.4 \\[1.3ex]
$I$              & 7930& $-22.74^{+ 0.07}_{- 0.08}$ & $ -1.31 ^{+ 0.01}_{- 0.01}$          &$  0.80^{+ 0.05}_{- 0.05}$ &          -15.74 \\[1.3ex]
2MASS   $H$      & 7754& $-23.44^{+ 0.07}_{- 0.08}$ & $ -1.29 ^{+ 0.01}_{- 0.01}$          &$  0.78^{+ 0.05}_{- 0.05}$ &         -16.21 \\[1.3ex]
2MASS   $J$      & 7834& $-23.21^{+ 0.08}_{- 0.07}$ & $ -1.30 ^{+ 0.01}_{- 0.01}$          &$  0.79^{+ 0.05}_{- 0.04}$ &         -16.04 \\[1.3ex]
2MASS \textit{Ks}& 7668& $-23.18^{+ 0.07}_{- 0.08}$ & $ -1.29 ^{+ 0.01}_{- 0.01}$          &$  0.78^{+ 0.05}_{- 0.04}$ &         -15.96 \\[1.3ex]
UKIDSS  $H$      & 8006& $-23.43^{+ 0.07}_{- 0.08}$ & $ -1.29 ^{+ 0.01}_{- 0.01}$          &$  0.78^{+ 0.04}_{- 0.04}$ &         -16.11 \\[1.3ex]
UKIDSS  $J$      & 7819& $-23.22^{+ 0.08}_{- 0.08}$ & $ -1.30 ^{+ 0.01}_{- 0.01}$          &$  0.78^{+ 0.05}_{- 0.05}$ &         -16.05 \\[1.3ex]
UKIDSS  $K$      & 7819& $-23.13^{+ 0.08}_{- 0.08}$ & $ -1.29 ^{+ 0.01}_{- 0.01}$          &$  0.78^{+ 0.05}_{- 0.04}$ &         -15.86 \\[1.3ex]
UKIDSS  $Y$      & 7839& $-23.06^{+ 0.07}_{- 0.08}$ & $ -1.30 ^{+ 0.01}_{- 0.01}$          &$  0.79^{+ 0.05}_{- 0.05}$ &         -15.95 \\[1.3ex]
UKIDSS  $Z$      & 8206& $-22.82^{+ 0.07}_{- 0.08}$ & $ -1.31 ^{+ 0.01}_{- 0.01}$          &$  0.79^{+ 0.05}_{- 0.05}$ &        -15.68 \\[1.3ex]
  $u$            & 4714& $-19.78^{+ 0.09}_{- 0.09}$ & $ -1.34 ^{+ 0.02}_{- 0.02}$          &$  0.80^{+ 0.07}_{- 0.06}$ &          -14.51 \\[1.3ex]
  $g$            & 7872& $-21.45^{+ 0.08}_{- 0.08}$ & $ -1.32 ^{+ 0.01}_{- 0.01}$          &$  0.81^{+ 0.05}_{- 0.05}$ &          -14.67 \\[1.3ex]
  $r$            & 8082& $-22.22^{+ 0.07}_{- 0.08}$ & $ -1.32 ^{+ 0.01}_{- 0.01}$          &$  0.80^{+ 0.05}_{- 0.05}$ &         -15.26 \\[1.3ex]
  $i$            & 8128& $-22.57^{+ 0.08}_{- 0.08}$ & $ -1.31 ^{+ 0.01}_{- 0.01}$          &$  0.80^{+ 0.05}_{- 0.05}$ &         -15.56 \\[1.3ex]
  $z$            & 7915& $-22.82^{+ 0.07}_{- 0.08}$ & $ -1.30 ^{+ 0.01}_{- 0.01}$          &$  0.80^{+ 0.05}_{- 0.05}$ &          -15.79 \\[1.3ex]
\bottomrule
\end{tabular}
\end{table}

%% file: tab_fits2sc.tex
\begin{table}
    \caption{Best fit parameters for the double Schechter function fit, inside $r_{200}$ at $z=0$. Only the $V$, $g$ and $r$ filters of the intermediate mass bin are presented, since they are the only cases where a double Schecther fit is preferred. Values and errors are the median and the $1\sigma$ percentile of the posterior parameter distribution. The columns refer to: (1) filter name, (2) number of galaxies selected, (3) knee magnitude (in mags), (4) slope and (5) normalization (in $h^3 \cMpc^{-3} \mathrm{mag}^{-1}$) of the bright Schechter function, (6) slope and (7) normalization of the faint Schechter function, (8) $\Delta \mathrm{BIC}$ (high values mean a stronger evidence for a double Schechter fit, see Section~\ref{sec_lumFunc}). A machine readable version is present in the online material, along with a plot of the posterior for each fit.}\label{tabDouble}
\begin{tabular}{ccccccccc}
\toprule
Filter   & N    & $M^*$                     & $\alpha_\mathrm{b}$         & $\phi_\mathrm{b}$          & $\alpha_\mathrm{f}$        & $\phi_\mathrm{f}$          & $\Delta\mathrm{BIC}$ \\
    (1) & (2) & (3) & (4) & (5) & (6) & (7) & (8) \\
\midrule
$V$      & 3220 & $-20.97^{+ 0.21}_{- 0.21}$ & $ -0.41^{+ 0.31}_{- 0.30}$ & $  1.76^{+ 0.33}_{- 0.34}$ & $ -1.39^{+ 0.04}_{- 0.06}$ & $  0.83^{+ 0.23}_{- 0.27}$ & 2.6 \\[1.3ex]
$g$      & 3205 & $-20.57^{+ 0.21}_{- 0.23}$ & $ -0.40^{+ 0.33}_{- 0.33}$ & $  1.80^{+ 0.33}_{- 0.33}$ & $ -1.39^{+ 0.04}_{- 0.07}$ & $  0.82^{+ 0.24}_{- 0.31}$ & 3.8  \\[1.3ex]
$r$      & 3288 & $-21.35^{+ 0.20}_{- 0.20}$ & $ -0.45^{+ 0.31}_{- 0.28}$ & $  1.73^{+ 0.33}_{- 0.34}$ & $ -1.39^{+ 0.04}_{- 0.06}$ & $  0.80^{+ 0.23}_{- 0.26}$ & 2.1  \\[1.3ex]

\bottomrule
\end{tabular}
\end{table}